\documentclass[conference]{IEEEtran}
\usepackage{etex}

\usepackage[font={sf,scriptsize}]{caption}
\usepackage[font={sf,scriptsize}]{subcaption}

\usepackage{makecell,cellspace}
\usepackage{graphicx}
\graphicspath{{figs/},{plots/}}
\usepackage{booktabs}
\usepackage{epstopdf}
\usepackage{url}
\usepackage{placeins}
\usepackage{amsmath,amssymb,amsfonts,amsthm}
\usepackage{mathtools,mathrsfs}
\usepackage{multirow}
\usepackage{rotating}
\usepackage{pbox}
\usepackage[normalem]{ulem}

\usepackage{inconsolata}
\usepackage{listings}

\newcommand{\subparagraph}{}
\usepackage[compact]{titlesec}
\titlespacing*{\section}{0pt}{6pt}{2pt}
\titlespacing*{\subsection}{0pt}{5pt}{1pt}
\titlespacing*{\subsubsection}{0pt}{5pt}{1pt}

\usepackage{etoolbox}

\makeatletter
\patchcmd{\ttlh@hang}{\parindent\z@}{\parindent\z@\leavevmode}{}{}
\patchcmd{\ttlh@hang}{\noindent}{}{}{}
\makeatother

\usepackage{cleveref}
\usepackage[utf8]{inputenc}
\crefname{section}{§}{§§}
\Crefname{section}{§}{§§}

\usepackage[10pt]{moresize}

\usepackage{enumitem}

\newcommand{\macs}[1]{{\noindent\small\textbf{\textsf{#1}}}}

\usepackage{scalerel,stackengine}
\stackMath
\newcommand\rwh[1]{%
	\savestack{\tmpbox}{\stretchto{%
			\scaleto{%
				\scalerel*[\widthof{\ensuremath{#1}}]{\kern-.6pt\bigwedge\kern-.6pt}%
				{\rule[-\textheight/2]{1ex}{\textheight}}%WIDTH-LIMITED BIG WEDGE
			}{\textheight}% 
		}{0.5ex}}%
	\stackon[1pt]{#1}{\tmpbox}%
}

\usepackage{amsmath,amssymb,amsfonts}
\usepackage{graphicx}
\usepackage{textcomp}
\usepackage{xcolor}
\def\BibTeX{{\rm B\kern-.05em{\sc i\kern-.025em b}\kern-.08em
    T\kern-.1667em\lower.7ex\hbox{E}\kern-.125emX}}

\usepackage{fontawesome}
\usepackage{pifont}
\usepackage{textcomp}

\usepackage{xcolor}
\usepackage{soul}

\usepackage{dblfloatfix}

\usepackage{mathrsfs}
\usepackage{amsmath}

\DeclareMathOperator*{\argmin}{arg\,min}

\usepackage{algorithm,algorithmicx,algpseudocode}

\usepackage{tikz}
\usetikzlibrary{tikzmark}

\usepackage{xpatch}
\expandafter\xpatchcmd
\csname pgfk@/tikz/every picture/.@cmd\endcsname
{\thepage}{\arabic{page}}{}{}

\tikzstyle{comment} = [draw, fill=blue!70, text=white, text width=3cm, minimum height=1cm, rounded corners, align=left, font=\scriptsize]
\tikzstyle{background_alg} = [draw, fill=blue!20, opacity=0.4, inner sep=4pt, rounded corners=2pt]

\usetikzlibrary{shapes}
\usetikzlibrary{plotmarks}
\usetikzlibrary{calc,fit}

\definecolor{vlgray}{rgb}{0.91 0.91 0.91}
\definecolor{ablack}{rgb}{0.2 0.2 0.2}

\usepackage[customcolors]{hf-tikz}
\hfsetbordercolor{white}
\hfsetfillcolor{vlgray}

\newcounter{highlight}

\newcounter{Ahighlight}

\definecolor{armygreen}{rgb}{0.05,0.75,0.0}

\newcommand{\myboxX}[1]{\tikz[baseline={(a.base)}]\node[draw=black,rounded corners=0.5ex,fill=ablack,inner sep=1pt](a){\textcolor{white}{#1}};}

\usetikzlibrary{matrix,shapes,arrows,fit,tikzmark}
\newcolumntype{.}{D{.}{.}{-1}}
\tikzset{   
        every picture/.style={remember picture,baseline},
        every node/.style={anchor=base,align=center,outer sep=-2pt},
        every path/.style={thick},
        }

\newcommand\marktopleftR[1]{%
    \tikz[overlay,remember picture] 
        \node (marker-#1-a) at (.3em,.3em) {};%
}
\newcommand\markbottomrightR[1]{%
    \tikz[overlay,remember picture] 
        \node (marker-#1-b) at (-.2em,.3em) {};%
    \tikz[overlay,remember picture,inner sep=3pt]
        \node[fill opacity=0.3,fill=red,rounded corners,fit=(marker-#1-a.north west) (marker-#1-b.south east)] {};%
}

\newcommand\marktopleftG[1]{%
    \tikz[overlay,remember picture] 
        \node (marker-#1-a) at (.3em,.3em) {};%
}
\newcommand\markbottomrightG[1]{%
    \tikz[overlay,remember picture] 
        \node (marker-#1-b) at (-.2em,.3em) {};%
    \tikz[overlay,remember picture,inner sep=3pt]
        \node[fill opacity=0.3,fill=armygreen,rounded corners,fit=(marker-#1-a.north west) (marker-#1-b.south east)] {};%
}

\newcommand\marktopleftGG[1]{%
    \tikz[overlay,remember picture] 
        \node (marker-#1-a) at (.0em,-0.35em) {};%
}
\newcommand\markbottomrightGG[1]{%
    \tikz[overlay,remember picture] 
        \node (marker-#1-b) at (.1em,1em) {};%
    \tikz[overlay,remember picture,inner sep=3pt]
        \node[fill opacity=0.3,fill=armygreen,rounded corners,fit=(marker-#1-a.north west) (marker-#1-b.south east)] {};%
}

\newcommand\marktopleftB[1]{%
    \tikz[overlay,remember picture] 
        \node (marker-#1-a) at (.3em,.3em) {};%
}
\newcommand\markbottomrightB[1]{%
    \tikz[overlay,remember picture] 
        \node (marker-#1-b) at (-.2em,.3em) {};%
    \tikz[overlay,remember picture,inner sep=3pt]
        \node[fill opacity=0.2,fill=blue,rounded corners,fit=(marker-#1-a.north west) (marker-#1-b.south east)] {};%
}

\begin{document}

\title{\vspace{-0.5em}A Modular Benchmarking Infrastructure for High-Performance and Reproducible Deep Learning\vspace{-0.25em}}

\author{\IEEEauthorblockN{Tal Ben-Nun, Maciej Besta, Simon Huber, Alexandros Nikolaos Ziogas, Daniel Peter, Torsten Hoefler}
\IEEEauthorblockA{\textit{Department of Computer Science, ETH Zurich
\vspace{-0.25em}
} \\
}
}

\maketitle

\begin{abstract}
We introduce Deep500: the first customizable benchmarking infrastructure that
enables fair comparison of the plethora of deep learning frameworks,
algorithms, libraries, and techniques. The key idea behind Deep500 is its
\emph{modular} design, where deep learning is factorized into four distinct
\emph{levels}: operators, network processing, training, and distributed
training. Our evaluation illustrates that Deep500 is \emph{customizable}
(enables combining and benchmarking different deep learning codes) and
\emph{fair} (uses carefully selected metrics).  Moreover, Deep500 is
\emph{fast} (incurs negligible overheads), \emph{verifiable} (offers
infrastructure to analyze correctness), and \emph{reproducible}.  Finally, as
the first distributed and reproducible benchmarking system for deep learning,
Deep500 provides software infrastructure to utilize the most powerful supercomputers for extreme-scale workloads.  \end{abstract}

\begin{IEEEkeywords}
Deep Learning, High-Performance Computing, Benchmarks, Distributed Deep Learning
\end{IEEEkeywords}

{\noindent\small\textbf{Deep500 Code:} \url{https://www.github.com/deep500/deep500}}

\section{Introduction}

Deep Learning (DL) has transformed the world and is now ubiquitous in areas
such as speech recognition, image classification, or autonomous
driving~\cite{ben2018demystifying}. Its central concept is a Deep Neural
Network (DNN), a structure modeled after the human brain. Thanks to rigorous
training, DNNs are able to solve various problems, previously deemed unsolvable.

Recent years saw an unprecedented growth in the number of approaches, schemes,
algorithms, applications, platforms, and frameworks for DL. First, DL
computations can aim at inference or training. Second, hardware platforms can
vary significantly, including CPUs, GPUs, or FPGAs. Third, operators can be
computed using different methods, e.g., im2col~\cite{im2col} or
Winograd~\cite{lavin2016fast} in convolutions.  Next, DL functionalities have
been deployed in a variety of frameworks, such as
TensorFlow~\cite{dong2017tensorlayer} or Caffe~\cite{jia2014caffe}. These
functionalities may incorporate many parallel and distributed optimizations,
such as data, model, and pipeline parallelism. Finally, DL workloads are
executed in wildly varying environments, such as mobile phones, multi-GPU
clusters, or large-scale supercomputers.

\begin{table}[b!]
\vspace{-2em}
\setlength{\tabcolsep}{0.3pt}
\renewcommand{\arraystretch}{0.8}
\centering
\scriptsize
\sf
\begin{tabular}{lccccccccccccc}
\toprule
\multirow{2}{*}{\textbf{System}} & \multicolumn{2}{c}{Operators} & \multicolumn{4}{c}{Networks} & \multicolumn{3}{c}{Training} & \multicolumn{4}{c}{Dist. Training} \\
\cmidrule(lr){2-3} \cmidrule(lr){4-7} \cmidrule(lr){8-10} \cmidrule(lr){11-14}
 & \textbf{Sta} & \textbf{Cus} & \textbf{Def} & \textbf{Eag} & \textbf{Com} & \textbf{Tra} & \textbf{Dat} & \textbf{Opt} & \textbf{Cus} & \textbf{~PS} & \textbf{Dec} & \textbf{Asy} & \textbf{Cus} \\
\midrule
\textbf{(L)} cuDNN & \marktopleftG{g1}\faThumbsOUp &\faThumbsDown&\marktopleftR{r1}\faThumbsDown&\faThumbsDown&\faThumbsDown&\faThumbsDown&\faThumbsDown&\faThumbsDown&\faThumbsDown&\faThumbsDown&\faThumbsDown&\faThumbsDown&\faThumbsDown  \\
\textbf{(L)} MKL-DNN & \faThumbsOUp &\faThumbsDown \markbottomrightG{g1}&\faThumbsDown&\faThumbsDown&\faThumbsDown&\faThumbsDown&\faThumbsDown&\faThumbsDown&\faThumbsDown&\faThumbsDown&\faThumbsDown&\faThumbsDown&\faThumbsDown \markbottomrightR{r1} \\
\midrule
\textbf{(F)} TensorFlow~\cite{abadi2016tensorflow} &\marktopleftG{g2} \faThumbsOUp & \faThumbsOUp & \marktopleftG{g3} \faThumbsOUp & \faThumbsOUp & \faThumbsUp & \faThumbsUp &  \marktopleftG{g4} \faThumbsOUp & UR & \faThumbsOUp &\marktopleftG{g5} \faThumbsOUp & \faThumbsUp & \faThumbsOUp & \faThumbsDown \\ 
\textbf{(F)} Caffe, Caffe2$^\text{\textdagger}$~\cite{jia2014caffe} &\faThumbsOUp&\faThumbsUp& \faThumbsOUp & \faThumbsDown  &\faThumbsDown&\faThumbsDown&\faThumbsDown&UR&\faThumbsDown&\faThumbsDown&\faThumbsOUp&\faThumbsDown&\faThumbsDown \\
\textbf{(F)} [Py]Torch$^\text{\textdagger}$~\cite{collobert2002torch, paszke2017automatic} &\faThumbsOUp&\faThumbsOUp&\faThumbsDown& \faThumbsOUp &\faThumbsDown&\faThumbsDown&\faThumbsDown&\faThumbsOUp&\faThumbsOUp&\faThumbsDown&\faThumbsOUp&\faThumbsDown&\faThumbsDown \\
\textbf{(F)} MXNet~\cite{chen2015mxnet} &\faThumbsOUp&\faThumbsUp&\faThumbsOUp & \faThumbsDown &\faThumbsDown&\faThumbsDown&\faThumbsOUp&UR&\faThumbsUp&\faThumbsOUp&\faThumbsDown&\faThumbsOUp&\faThumbsDown \\
\textbf{(F)} CNTK~\cite{yu2014introduction} &\faThumbsOUp&\faThumbsUp& \faThumbsOUp & \faThumbsDown &\faThumbsDown&\faThumbsDown&\faThumbsOUp&UR&\faThumbsUp&\faThumbsDown&\faThumbsOUp&\faThumbsDown&\faThumbsDown \markbottomrightG{g5} \\
\textbf{(F)} Theano~\cite{bergstra2011theano} &\faThumbsOUp&\faThumbsOUp& \faThumbsOUp & \faThumbsOUp &\faThumbsOUp&\faThumbsOUp&\faThumbsDown&\faThumbsUp&\faThumbsDown&\marktopleftR{r2}\faThumbsDown&\faThumbsDown&\faThumbsDown&\faThumbsDown \markbottomrightR{r2}\\
\textbf{(F)} Chainer[MN]~\cite{tokui2015chainer} &\faThumbsOUp&\faThumbsOUp& \faThumbsDown & \faThumbsOUp &\faThumbsDown&\faThumbsDown\markbottomrightG{g3} &\faThumbsOUp &\faThumbsUp&\faThumbsOUp \markbottomrightG{g4}&\marktopleftG{g5}\faThumbsDown&\faThumbsOUp&\faThumbsDown&\faThumbsDown\markbottomrightG{g5} \\
\textbf{(F)} Darknet~\cite{darknet13} &\faThumbsOUp&\faThumbsDown&\marktopleftB{y1}\faThumbsOUp & \faThumbsDown  &\faThumbsDown&\faThumbsDown& \marktopleftB{b2} \faThumbsDown&\faThumbsUp&\faThumbsDown \markbottomrightB{b2} & \marktopleftR{r4} \faThumbsDown&\faThumbsDown&\faThumbsDown&\faThumbsDown \markbottomrightR{r4} \\
\textbf{(F)} DL4j~\cite{team2016deeplearning4j} &\faThumbsOUp&\faThumbsOUp& \faThumbsOUp &  \faThumbsDown &\faThumbsDown&\faThumbsDown& \marktopleftG{g7} \faThumbsOUp&UR&\faThumbsUp \markbottomrightG{g7} & \marktopleftG{g9} \faThumbsOUp&\faThumbsDown&\faThumbsDown&\faThumbsDown \markbottomrightG{g9} \\
\textbf{(F)} DSSTNE &\faThumbsOUp&\faThumbsDown& \faThumbsOUp &\faThumbsDown   &\faThumbsDown&\faThumbsDown& \marktopleftB{b3} \faThumbsDown&UR&\faThumbsDown \markbottomrightB{b3} & \marktopleftR{r5} \faThumbsDown  &\faThumbsDown&\faThumbsDown&\faThumbsDown \markbottomrightR{r5} \\
\textbf{(F)} PaddlePaddle &\faThumbsOUp&\faThumbsOUp& \faThumbsOUp & \faThumbsDown  &\faThumbsDown&\faThumbsUp \markbottomrightB{y1} & \marktopleftG{g8} \faThumbsUp&UR&\faThumbsUp \markbottomrightG{g8} & \marktopleftG{g10} \faThumbsOUp&\faThumbsDown&\faThumbsOUp&\faThumbsUp \markbottomrightG{g10} \\ % Reason for semi-thumbs up on Customizable Distributed Training: Using an IR transpiler to distribute computations (e.g., model parallelism), and allowing control between async, sync, and model averaging dist sgd. Plus, elastic training.
\textbf{(F)} TVM \cite{tvm} &\faThumbsOUp&\faThumbsOUp \markbottomrightG{g2}& \marktopleftG{g6} \faThumbsOUp & \faThumbsDown  &\faThumbsOUp&\faThumbsOUp \markbottomrightG{g6} & \marktopleftR{r3} \faThumbsDown&\faThumbsDown&\faThumbsDown \markbottomrightR{r3} & \marktopleftR{r6} \faThumbsDown&\faThumbsDown&\faThumbsDown&\faThumbsDown \markbottomrightR{r6} \\
\midrule 
\textbf{(E)} Keras~\cite{chollet2015others} &\marktopleftG{g11}\faThumbsOUp&\faThumbsDown\markbottomrightG{g11}&\marktopleftR{r11}\faThumbsDown&\faThumbsDown&\faThumbsDown&\faThumbsDown&\marktopleftB{y11}\faThumbsDown&UR&\faThumbsUp\markbottomrightB{y11}&\marktopleftR{r12}\faThumbsDown&\faThumbsDown&\faThumbsDown&\faThumbsDown\markbottomrightR{r12} \\
\textbf{(E)} Horovod~\cite{sergeev2018horovod} &\marktopleftR{r14}\faThumbsDown&\faThumbsDown\markbottomrightR{r14}&\faThumbsDown&\faThumbsDown&\faThumbsDown&\faThumbsDown&\marktopleftR{r15}\faThumbsDown&\faThumbsDown&\faThumbsDown\markbottomrightR{r15}&\marktopleftG{g12}\faThumbsDown&\faThumbsOUp&\faThumbsDown&\faThumbsDown \\
\textbf{(E)} TensorLayer~\cite{dong2017tensorlayer} &\marktopleftG{g13}\faThumbsOUp&\faThumbsDown&\faThumbsDown&\faThumbsDown&\faThumbsDown&\faThumbsDown&\marktopleftB{y12}\faThumbsDown&UR&\faThumbsDown&\faThumbsDown&\faThumbsUp&\faThumbsDown& \faThumbsDown\markbottomrightG{g12} \\
\textbf{(E)} Lasagne &\faThumbsUp&\faThumbsOUp&\faThumbsDown&\faThumbsDown&\faThumbsDown&\faThumbsDown&\faThumbsDown&UR&\faThumbsUp&\marktopleftR{r13}\faThumbsDown&\faThumbsDown& \faThumbsDown&\faThumbsDown \\
\textbf{(E)} TFLearn~\cite{tflearn} &\faThumbsUp&\faThumbsDown\markbottomrightG{g13}&\faThumbsDown&\faThumbsDown&\faThumbsDown&\faThumbsDown\markbottomrightR{r11}&\faThumbsOUp&\faThumbsDown&\faThumbsDown\markbottomrightB{y12}&\faThumbsDown&\faThumbsDown&\faThumbsDown& \faThumbsDown\markbottomrightR{r13} \\
\midrule
\makecell[l]{\textbf{Integration within}\\\textbf{Deep500 [This work]}} & \marktopleftGG{g999}\faThumbsOUp$^{*\ddagger}$ & \faThumbsOUp$^{*}$ & 
\faThumbsOUp$^{*\ddagger}$ & \faThumbsOUp$^{*\ddagger}$ &\faThumbsOUp &\faThumbsOUp$^{*}$&
\faThumbsOUp$^{*\ddagger}$ &\faThumbsOUp$^{*\ddagger}$ &\faThumbsOUp$^{*}$ &
\faThumbsOUp$^{*\ddagger}$ &\faThumbsOUp$^{*\ddagger}$ &\faThumbsOUp$^{*\ddagger}$&\faThumbsOUp$^{*}$
\markbottomrightGG{g999} \\
\bottomrule
\end{tabular}
\vspace{-0.5em}
\caption{
\textbf{An overview of DL frameworks, related systems that can be integrated
within Deep500, and the advantages of such integration}.
Each column is a specific feature/functionality; they are explained in
more detail in Background (\cref{sec:background}).
\textbf{Sta}: Standard Operators,
\textbf{Cus}: Customizable (without full recompilation),
\textbf{Def}: Deferred Execution Mode,
\textbf{Eag}: Eager Execution Mode (also called ``define-by-run''),
\textbf{Com}: Network Compilation,
\textbf{Tra}: Transformable,
\textbf{Dat}: Dataset Network Integration,
\textbf{Opt}: Standard Optimizers,
\textbf{PS}: Parameter Server,
\textbf{Dec}: Decentralized,
\textbf{Asy}: Asynchronous SGD,
\textbf{UR}: Update Rule Optimizers,
\faThumbsOUp, \faThumbsUp, \faThumbsDown: A given system does offer a given feature, offers a given feature
in a limited way, or does not offer it.
\textbf{(L)}: a library, 
\textbf{(F)}: a framework,
\textbf{(E)}: a frontend.
$^\text{\textdagger}$Caffe2 and PyTorch exist as a single repository of code, but execute as two separate frameworks.
$^{*}$ Deep500 provides an isolated modular abstraction of a given feature.
$^{\ddagger}$Deep500 provides a reference implementation.
Native system support for a category of features: \colorbox{red!30}{none},
\colorbox{blue!30}{partial}, \colorbox{armygreen!30}{full}.  }
\vspace{-1em}
\label{tab:intro-frameworks}
\end{table}

This richness of the DL domain raises an important question: \textbf{How can one ensure a leveled, fair ground for comparison,
competition, and benchmarking in Deep Learning?}
The key issue is that, due to the complex nature of DL workloads, there is no single metric by which one DNN or hardware is objectively better than another on all counts. This is an open question and there are multiple proposed metrics (e.g., throughput, time-to-solution), especially for hardware ranking for distributed DL. Thus, it is necessary to design an abstraction that supports these and future potential metrics for DL evaluation. Several benchmarks~\cite{coleman2017dawnbench,mlp,tbd,fathom,dlcookbook,kaggle,imagenet,mlp} have been designed for DL, but they are either specialized for a given aspect (e.g., single-layer performance) or perform black-box tests, which hinder verifiability and reproducibility. Since DL is converging in terms of procedures, it is possible to design a white-box abstraction that covers key functionalities of the problem, enabling \textit{arbitrary} metric measurement and full integration of the different software stacks (see Table~\ref{tab:intro-frameworks}) for benchmarking.

We propose \textbf{Deep500}: a white-box benchmarking infrastructure that enables fair analysis and comparison of diverse DL workloads and algorithms.
Deep500 is based on the following \textbf{five pillars}: \ding{182}
\textbf{Customizability}, \ding{183} \textbf{Metrics}, \ding{184}
\textbf{Performance}, \ding{185} \textbf{Validation}, and \ding{186}
\textbf{Reproducibility}.
\ding{182} ``Customizability'' indicates that Deep500 enables benchmarking of arbitrary
combinations of DL elements, such as various frameworks running on different
platforms, and executing custom algorithms. To achieve this, we design
Deep500 to be a \emph{meta-framework} that can be straightforwardly extended to
benchmark any DL code. Table~\ref{tab:intro-frameworks} illustrates how
various DL frameworks, libraries, and frontends can be integrated in Deep500 to
enable easier and faster DL programming.
\ding{183} ``Metrics'' indicates that Deep500 embraces a complex nature of DL that, unlike
benchmarks such as Top500~\cite{dongarra1994top500}, makes a single number such as FLOPS an insufficient
measure. To this end, we propose metrics that consider the accuracy-related
aspects of DL (e.g., time required to ensure a specific test-set accuracy) and
performance-related issues (e.g., communication volume).
\ding{184} ``Performance'' means that Deep500 is the first DL benchmarking infrastructure
that can be integrated with parallel and distributed DL codes.
\ding{185} ``Validation'' indicates that Deep500 provides infrastructure to ensure
correctness of aspects such as convergence.
Finally, Deep500 embraces \ding{186} ``Reproducibility'' as specified
in recent HPC initiatives~\cite{hoefler2015scientific} to help
developing reproducible DL codes.

Table~\ref{tab:intro-benchmarks} compares Deep500 to other benchmarking
infrastructures with respect to the offered functionalities. Deep500 is the
only system that focuses on performance, accuracy, and convergence, while
simultaneously offering a wide spectrum of metrics and criteria for
benchmarking, enabling customizability of design, and considering a diversity
of workloads.

Towards these goals, we contribute:

\begin{itemize}[noitemsep, leftmargin=1em]
\item the identification and analysis of challenges in high-performance
reproducible benchmarking of deep learning,
\item the design and implementation of the \textbf{Deep500 Benchmark}, a
\emph{meta-framework} for customizable, fast, validatable, and reproducible
benchmarking of \emph{extreme-scale} DL frameworks, applications, algorithms,
and techniques,
\item extensive evaluation, illustrating that Deep500 \ding{182} incurs
negligible ($<$1\%) overheads of benchmarking on top of tested systems, and
\ding{183} vastly reduces development efforts to integrate and benchmark
various elements of deep learning.
\end{itemize}

\captionsetup[table]{font={scriptsize,sf},labelfont={scriptsize,sf}}

\begin{table*}[t!]
\vspace{-1em}
\setlength{\tabcolsep}{1pt}
\renewcommand{\arraystretch}{0.9}
\centering
\scriptsize
\sf
\begin{tabular}{lccccccccccccccccccccccccp{0.85in}}
\toprule
\multirow{2}{*}{\textbf{Benchmark}} & \multicolumn{3}{c}{Focus} & \multicolumn{9}{c}{Metrics} & \multicolumn{3}{c}{Criteria} & \multicolumn{3}{c}{Customizability} &  \multicolumn{6}{c}{DL Workloads} & Remarks \\
\cmidrule(lr){2-4}\cmidrule(lr){5-13}\cmidrule(lr){14-16}\cmidrule(lr){17-19}\cmidrule(lr){20-25}
&\textbf{Perf}& \textbf{Con} & \textbf{Acc}&
\textbf{Tim}&\textbf{Cos}&\textbf{Ene}&\textbf{Util}&\textbf{Mem}&\textbf{Tput}&\textbf{Brk}&\textbf{Sca}&\textbf{Com}&
\textbf{TTA}&\textbf{FTA}&\textbf{Lat}&%
\textbf{Clo}&\textbf{Ope}&\textbf{Inf}&%
\textbf{Ops}&\textbf{Img}&\textbf{Obj}&\textbf{Spe}&\textbf{Txt}&\textbf{RL}&\\
\midrule
DeepBench~\cite{deepbench} & \faThumbsOUp &\faThumbsDown&\faThumbsDown & \faThumbsOUp &\faThumbsDown&\faThumbsDown&\faThumbsDown&\faThumbsDown&\faThumbsDown&\faThumbsDown&\faThumbsDown&\faThumbsDown&\faThumbsDown&\faThumbsDown&\faThumbsOUp&\faThumbsOUp&\faThumbsDown&\faThumbsOUp&\faThumbsOUp&\faThumbsDown&\faThumbsDown&\faThumbsDown&\faThumbsDown&\faThumbsDown&Ops: Conv., GEMM, RNN, Allreduce\\
TBD~\cite{tbd} & \faThumbsOUp &\faThumbsDown&\faThumbsDown &\faThumbsDown&\faThumbsDown&\faThumbsDown&\faThumbsOUp&\faThumbsOUp&\faThumbsOUp&\faThumbsDown&\faThumbsDown&\faThumbsDown&\faThumbsDown&\faThumbsDown&\faThumbsOUp&\faThumbsOUp&\faThumbsDown&\faThumbsDown&\faThumbsDown&\faThumbsOUp&\faThumbsOUp&\faThumbsOUp&\faThumbsOUp&\faThumbsOUp & +GANs \\
Fathom~\cite{fathom} & \faThumbsOUp & \faThumbsDown&\faThumbsDown & \faThumbsOUp &\faThumbsDown&\faThumbsDown&\faThumbsDown&\faThumbsDown&\faThumbsOUp&\faThumbsOUp&\faThumbsOUp&\faThumbsDown&\faThumbsDown&\faThumbsDown&\faThumbsOUp&\faThumbsOUp&\faThumbsDown&\faThumbsDown&\faThumbsDown&\faThumbsOUp&\faThumbsDown&\faThumbsOUp&\faThumbsOUp&\faThumbsOUp& +Auto-encoders\\
DLBS~\cite{dlcookbook} & \faThumbsOUp&\faThumbsDown&\faThumbsDown& \faThumbsOUp&\faThumbsDown&\faThumbsDown&\faThumbsDown&\faThumbsDown&\faThumbsOUp&\faThumbsDown&\faThumbsDown&\faThumbsDown& \faThumbsDown&\faThumbsDown&\faThumbsDown&\faThumbsOUp&\faThumbsDown&\faThumbsOUp& \faThumbsDown&\faThumbsOUp&\faThumbsDown&\faThumbsDown&\faThumbsDown&\faThumbsDown&\\
DAWNBench~\cite{coleman2017dawnbench} &\faThumbsOUp & \faThumbsOUp &\faThumbsDown & \faThumbsOUp & \faThumbsOUp&\faThumbsDown&\faThumbsDown&\faThumbsDown&\faThumbsDown&\faThumbsDown&\faThumbsDown&\faThumbsDown&\faThumbsOUp&\faThumbsDown&\faThumbsOUp&\faThumbsDown&\faThumbsOUp&\faThumbsDown&\faThumbsDown&\faThumbsOUp&\faThumbsDown&\faThumbsDown&\faThumbsOUp&\faThumbsDown& \\
Kaggle~\cite{kaggle} &\faThumbsDown &\faThumbsDown& \faThumbsOUp &\faThumbsDown&\faThumbsDown&\faThumbsDown&\faThumbsDown&\faThumbsDown&\faThumbsDown&\faThumbsDown&\faThumbsDown&\faThumbsDown&\faThumbsDown&\faThumbsOUp&\faThumbsDown&\faThumbsDown&\faThumbsOUp&\faThumbsDown&\faThumbsDown&\faThumbsOUp&\faThumbsOUp&\faThumbsOUp&\faThumbsOUp&\faThumbsOUp&Varying workloads\\
ImageNet \cite{imagenet} &\faThumbsDown &\faThumbsDown&\faThumbsOUp&\faThumbsDown&\faThumbsDown&\faThumbsDown&\faThumbsDown&\faThumbsDown&\faThumbsDown&\faThumbsDown&\faThumbsDown&\faThumbsDown&\faThumbsDown&\faThumbsOUp&\faThumbsDown&\faThumbsDown&\faThumbsOUp&\faThumbsDown&\faThumbsDown&\faThumbsOUp&\faThumbsOUp&\faThumbsDown&\faThumbsDown&\faThumbsDown&\\
MLPerf~\cite{mlp} & \faThumbsOUp&\faThumbsOUp&\faThumbsOUp& \faThumbsOUp&\faThumbsOUp&\faThumbsOUp&\faThumbsDown&\faThumbsDown&\faThumbsDown&\faThumbsDown&\faThumbsDown&\faThumbsDown&\faThumbsOUp&\faThumbsOUp&\faThumbsOUp&\faThumbsOUp&\faThumbsOUp&\faThumbsDown&\faThumbsDown&\faThumbsOUp&\faThumbsOUp&\faThumbsOUp&\faThumbsOUp&\faThumbsOUp&\\
\midrule
\textbf{Deep500} & \faThumbsOUp & \faThumbsOUp & \faThumbsOUp &\faThumbsOUp&\faThumbsOUp&\faThumbsUp&\faThumbsOUp&\faThumbsUp&\faThumbsOUp&\faThumbsOUp&\faThumbsOUp&\faThumbsOUp&\faThumbsOUp&\faThumbsOUp&\faThumbsOUp&\faThumbsOUp&\faThumbsOUp&\faThumbsOUp&\faThumbsOUp&\faThumbsOUp&\faThumbsUp&\faThumbsUp& \faThumbsUp&\faThumbsUp&  \\
\bottomrule
\end{tabular}
\vspace{-0.5em}
\caption{
\textbf{An overview of available DL benchmarks}, focusing on the offered functionalities.
\textbf{Perf}: Performance,
\textbf{Con}: Convergence,
\textbf{Acc}: Accuracy,
\textbf{Tim}: Time,
\textbf{Cos}: Cost,
\textbf{Ene}: Energy,
\textbf{Util}: Utilization,
\textbf{Mem}: Memory Footprint,
\textbf{Tput}: Throughput (Samples per Second),
\textbf{Brk}: Timing Breakdown,
\textbf{Sca}: Strong Scaling,
\textbf{Com}: Communication and Load Balancing,
\textbf{TTA}: Time to Accuracy,
\textbf{FTA}: Final Test Accuracy,
\textbf{Lat}: Latency (Inference),
\textbf{Clo}: Closed (Fixed) Model Contests,
\textbf{Ope}: Open Model Contests,
\textbf{Inf}: Fixed Infrastructure for Benchmarking,
\textbf{Ops}: Operator Benchmarks,
\textbf{Img}: Image Processing,
\textbf{Obj}: Object Detection and Localization,
\textbf{Spe}: Speech Recognition,
\textbf{Txt}: Text Processing and Machine Translation,
\textbf{RL}: Reinforcement Learning Problems,
\faThumbsOUp: A given benchmark does offer the feature. \faThumbsUp: Planned benchmark feature. \faThumbsDown: A given benchmark does not offer the feature.
}
\vspace{-1.5em}
\label{tab:intro-benchmarks}
\end{table*}

\section{Background}
\label{sec:background}

\begin{figure}
	\centering
	\includegraphics[width=0.8\columnwidth]{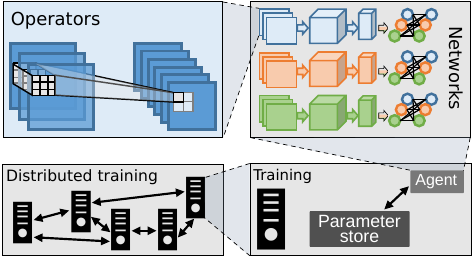}
  \vspace{0.0em}
	\caption{Components in Deep Learning}
	\label{fig:dl}
\end{figure}

We start with describing background on deep learning.

\subsection{Deep Learning}
\label{sec:dl}

We focus on \emph{Deep Learning} (DL): a subclass of Machine Learning (ML) that
uses Deep Neural Networks (DNNs)~\cite{ben2018demystifying} for approximating
certain complex functions.  In this paper, 
we mostly discuss \emph{supervised learning}, but Deep500 can be used to benchmark workloads for other tasks, such as unsupervised and reinforcement learning. A DNN is
first \emph{trained}: it is provided with various input data samples in a
randomized order to minimize the difference (\emph{loss}) between the obtained
and the desired outcome (this difference is computed with some loss function
$\ell$).  After training, a DNN is used to \emph{infer} outcomes for given
inputs.

Intuitively, a DNN is a composition of multiple functions called
\textit{operators}. Operators can range from fully-connected neural networks,
through multi-dimensional convolution, to recurrent operators that maintain state. The process of evaluating a given operator for a given input
data (referred to as a \emph{sample}) is called \textit{inference}.
These operators are organized as a Directed Acyclic Graph (DAG)
(Fig.~\ref{fig:dl}, top right). 

Formally, for an input dataset $S \subset \left(X \times Y\right) \sim
\mathcal{D}$ of labeled samples (sampled from a data distribution
$\mathcal{D}$), and a parametric model $f:X\rightarrow Y$ (denoted by
$f(w,x)$), the goal is to minimize the \emph{expected loss over the dataset},
i.e., find a \emph{minimizing set of parameters} $w^*=\argmin_w
\mathbb{E}_{(x,y)}\left[\ell\left(f(w,x), y\right)\right]$, where $\ell$ is
a certain norm function for assessing difference. 

\macs{Training}
To minimize the expected loss, we use algorithms such as Stochastic Gradient
Descent (SGD)~\cite{sgd} for training. In SGD, dataset elements are sampled at
random in \textit{minibatches} (data portions) of size~$B$; usually $16 \le B \le
64$k~\cite{batch64k}. In training, one iterates over the whole dataset (one
such loop iteration is called an \textit{epoch}) multiple times, and modifies
minimizing parameters $w^{(t)}$ at iteration~$t$ according to the \emph{average}
gradient and possibly historical values of $w$. Algorithm~\ref{alg:sgd} depicts
such an SGD optimizer with a weight update rule~$U$.

\setlength{\textfloatsep}{0.5em}
\begin{algorithm}[h]
\footnotesize
	\begin{algorithmic}[1]
		\For{$t = 0$ \textbf{to} $\frac{|S|}{B}\cdot \#epochs$} \Comment $|S|$: input dataset size
		\State $\vec{x},\vec{y}\leftarrow$ Sample $B$ elements from $S$ \label{alg:sgd:readsamples} \Comment $\vec{x},\vec{y}$: samples of input data
		\State $w_{mb}\leftarrow w^{(t)}$       \label{alg:sgd:readweights}\Comment Load minimizing parameters from iteration $t$
		\State $\vec{z} \leftarrow \ell(w_{mb}, \vec{x}, \vec{y})$ \label{alg:sgd:fwd} \Comment Inference; $\vec{z}$ is a full minibatch.
		\State $g_{mb}\leftarrow\frac{1}{B}\sum_{i=0}^{B}\nabla\ell(w_{mb}, \vec{z}_i)$ \label{alg:sgd:glocal} \label{alg:sgd:bwd} \Comment Backpropagation; $\vec{z}_i$ is a sample.
		\State $\Delta w\leftarrow U(g_{mb},w^{(0,\dots,t)}, t)$ \label{alg:sgd:accgrad} \Comment Apply an update rule
		\State $w^{(t+1)}\leftarrow w_{mb}+\Delta w$ \label{alg:sgd:writeweights} \Comment Store updated minimizing parameters 
		\EndFor
	\end{algorithmic}
	\caption{Minibatch Stochastic Gradient Descent~\cite{sgd}}
	\label{alg:sgd}
\end{algorithm}
%NOTE: 3-step optimizer modifies every iteration of the loop, modifies w_mb before inference, and replaces lines 6-7.

\macs{Distributed Training}
When distributing training among compute nodes, it is common to use data parallelism, i.e.,
partitioning across minibatches. The gradient average (Algorithm~\ref{alg:sgd},
line~\ref{alg:sgd:bwd}), necessary for descent, becomes a parallel reduction that
is performed collectively (\textit{allreduce} in MPI nomenclature).
Data-parallel distributed training can be implemented in one of two general
approaches: \textit{decentralized}, using an allreduce operation; or
\textit{centralized}, where a (possibly sharded) ``parameter server'' (PS) governs
optimization~\cite{dean2012large} by receiving individual gradients and
broadcasting back new parameters (Algorithm~\ref{alg:sgd}, lines
\ref{alg:sgd:readweights} and \ref{alg:sgd:bwd}--\ref{alg:sgd:writeweights}).
Deep500 enables all these distributed schemes while vastly reducing
development effort.

\subsection{Frameworks}
\label{sec:back-fr}

Many frameworks for training and inference exist; see Table~\ref{tab:intro-frameworks}. According to
GitHub~\cite{dlranking}, the three most popular DL frameworks are
TensorFlow~\cite{abadi2016tensorflow}, Caffe~\cite{jia2014caffe}, and
PyTorch~\cite{collobert2002torch}. We use these three frameworks (except for
Caffe, which we replace with Caffe2, an improved Caffe version written by
the same authors) as use cases to demonstrate the flexibility of Deep500. 
Now, the frameworks can differ vastly, ranging from how operators are
implemented and how extensible (\textit{Customizable}) the frameworks are; to how DNNs are evaluated
and trained. Some frameworks compute operators on-the-fly, i.e., as they are called (\textit{Eager Execution}). Others
construct a graph in advance (\textit{Deferred Execution}) and modify it for high-level optimizations (\textit{Transformable}), ahead-of-time \textit{Network Compilation}, or employ \textit{Dataset Integration} by adding data loading operators to the graph, enabling automatic pipelining of samples to accelerators and distributed storage integration. As for training, most frameworks support the aforementioned \textit{Update Rule} SGD, but some also provide other optimizers, or enable implementing arbitrary algorithms.

A complete DL framework provides:
\ding{182} operator representation and implementations,
\ding{183} network definition (connections between operators), \ding{184}
schemes for loading datasets and for data augmentation (i.e., increasing
variation in samples by perturbing data), \ding{185} inference and gradient
computation, \ding{186} stochastic optimization (training), and \ding{187}
distributed optimization and communication infrastructure.
Elements \ding{182}--\ding{183} are partially standardized by initiatives such
as ONNX~\cite{exchange2018onnx} and NNEF~\cite{nnef} that define portable (up
to framework limitations) file formats. Other elements are not standardized at
all. Moreover, frameworks do not provide standardized metrics, such as accuracy
and performance, which are absolutely necessary for scientific computing
purposes, high-performance computing, and reproducibility.
Table~\ref{tab:intro-frameworks} illustrates the various DL frameworks,
libraries, and frontends, and how Deep500 integrates them through its 
meta-framework design. These systems can then be analyzed using a set 
of carefully selected metrics.

\subsection{Benchmarks}

There exist preliminary efforts to benchmark DL and general ML.
Table~\ref{tab:intro-benchmarks} analyzes the functionalities of these efforts.
In general, only Deep500 focuses on performance,
 accuracy, \emph{and}
convergence, considering the five challenges of large-scale DL
benchmarking discussed in the introduction.

\subsection{Data Model and Format}
\label{sec:back-onnx}

We use the Open Neural Network Exchange (ONNX)~\cite{exchange2018onnx} format
to store DNNs reproducibly. ONNX provides a binary file format
capable of describing an arbitrary DAG
and standardizes a list of 118 common operators (as of version 1.3.0) used in
DL and in general ML. Many popular frameworks provide and are actively
improving interoperability with ONNX.
Thus, we select ONNX as the basis of data format for Deep500. To use
ONNX for reproducible training and to enable extensibility, we augment the
ONNX with additional built-in operators and with support for
user-defined operators.

\section{Benchmarking Deep Learning: Challenges}
\label{sec:challenges}

We analyze challenges in benchmarking large-scale DL.

\subsection{Motivation}
\label{sec:motivation}

We first describe motivating example use cases.

\noindent
\myboxX{Use Case 1}
We observe that different frameworks come with significant differences between
basic functionalities. For example, we present the initialization of 2D
convolution in TensorFlow and CNTK in
Listings~\ref{lst:tf_motivate}--\ref{lst:cntk_motivate}. TensorFlow uses 19
parameters while CNTK needs 10. Thus, {comparing fairly both frameworks in
metrics such as runtime or accuracy is unclear} (``Which parameters correspond
to each other?''), and for some operators \textit{impossible} due to
implementation differences. Among the factors that vary between frameworks is
data layout, which is unclear in the CNTK example.
For the case of performance, we use the Adam SGD optimizer~\cite{adam} as a
second example. As TensorFlow provides general tensor operators (using the
Eigen linear algebra library), it requires sequentially executing several short
operations on the GPU (e.g. subtraction, division) to compute the optimizer
update. Conversely, Caffe2 implements a specific ``Adam'' operator that
performs the entire update using a single GPU kernel, drastically reducing
invocation and GPU scheduling overheads. 
This \textit{operation fusion} in Caffe2 is a common optimization technique, and Deep500 enables straightforward comparison of such TensorFlow and Caffe2 instances, both in an isolated environment and as an integrated part of training over existing datasets.
In general, \emph{this use case calls for an
infrastructure that enables straightforward invocation of existing and custom
implementations, in order to enable simple, maintainable comparison and
benchmarking.}

\begin{table}
  \noindent\begin{minipage}[t]{.48\columnwidth}
    \begin{lstlisting}[basicstyle=\tt\tiny, caption=\textbf{TensorFlow: 19 parameters} to init 2D convolution., label=lst:tf_motivate]
tf.layers.conv2d(
  inputs, filters,
  kernel_size, strides, padding,
  data_format, dilation_rate,
  activation, use_bias,
  kernel_initializer, bias_initializer,
  kernel_constraint, bias_constraint,
  kernel_regularizer,
  activity_regularizer,
  trainable, name, reuse
)
    \end{lstlisting}
  \end{minipage}\hfill
  \begin{minipage}[t]{.48\columnwidth}
    \begin{lstlisting}[basicstyle=\tt\tiny, caption=\textbf{CNTK: 10 parameters} to init 2D convolution., label=lst:cntk_motivate]
cntk.layers.Convolution2D(
  filter_shape, num_filters,
  activation, init,
  pad, strides, bias, init_bias,
  reduction_rank, name
)
    \end{lstlisting}
  \end{minipage}
\end{table}

\noindent
\myboxX{Use Case 2}
Constructing a complex DNN may be a significant time investment. Yet, today's
DL frameworks do not enable a straightforward use of a network developed in a
different framework. For example, networks designed in TensorFlow cannot easily
be used in Caffe2 (e.g., due to the aforementioned differences in operators).
\emph{One would welcome a system that facilitates porting between different DNN
formats, in order to develop DNN-related techniques independently, as well as reuse networks across frameworks.}

\noindent
\myboxX{Use Case 3}
No single framework provides all the required functionalities. Thus, one may be
interested in extending a selected framework. Unfortunately, this is usually
difficult and time-consuming. For example, implementing a second-order
optimization, such as Stochastic L-BFGS~\cite{moritz16slbfgs}, requires a
training loop that is vastly different than that in Algorithm~\ref{alg:sgd}, which is the
basis of many frameworks.  Now, while some frameworks (e.g., TensorFlow) enable
the creation of custom training loops (e.g., using optimized tensor
operations), the CNTK \texttt{Learner} extension does not enable this
straightforwardly.
\emph{An infrastructure for combining the best of different DL frameworks
would be advantageous in such cases.}

\noindent
\myboxX{Use Case 4}
Many DL tools are distributed. In such cases,
\emph{to ensure scalability and high performance, one needs
a system that can benchmark and analyze aspects related to
distributed processing, such as the amount of communicated data.}

\noindent
\myboxX{Others}
There are many other situations requiring a standard benchmarking platform for
DL. They can be pictured by the following example questions: ``What is the
reduction in communication over the network, when a certain compression scheme
is applied in training?'', ``How to illustrate performance and power advantages
of using a novel ASIC for a particular class of DL workloads?'', ``How fast and
accurate is a certain provably optimal operator?'', ``What is the advantage of
using FPGAs for DL training?'',
``For a given DL workload, which one of the available machines will perform best?''.

\subsection{\textbf{Challenge 1: Customizability}}

The first challenge emerging from the above examples is \emph{customizability}:
the ability to seamlessly and effortlessly combine different features of
different frameworks and still be able to provide fair analysis of their
performance and accuracy.

\textbf{Deep500 enables customizability and interoperability with various DL
codes through its \emph{meta-framework} design.} By incorporating both Python
and C/C++ capabilities, we provide an infrastructure that can be
straightforwardly extended to virtually any DL framework or arbitrary operator
code.

\subsection{\textbf{Challenge 2: Metrics}}

Another challenge lies in a proper selection of metrics. On one hand, some
metrics may simply be too detailed, for example the number of cache misses in
2D convolution implemented in TensorFlow or Caffe2. Due to the sheer
complexity of such frameworks, this metric would probably not provide useful
insights in potential performance regressions. On the other hand, other metrics
may be too generic, for example simple runtime does not offer any meaningful
details and does not relate to accuracy. Thus, one must select
metrics that find the right balance between accuracy and genericness.

\textbf{In Deep500, we offer carefully selected metrics, considering
performance, correctness, and convergence in shared- as well as distributed-memory
environments.} 
\ding{182} Some metrics can test the performance of both the whole
computation and fine-grained elements, for example \emph{latency} or
\emph{overhead}. \ding{183} Others, such as \emph{accuracy} or \emph{bias},
assess the quality of a given algorithm, its convergence, and its
generalization towards previously-unseen data. \ding{184} We also combine
performance and accuracy (\emph{time-to-accuracy}) to analyze the tradeoffs.
\ding{185} Finally, we propose metrics for the distributed
part of DL codes: \emph{communication volume} and \emph{I/O
latency}.

\subsection{\textbf{Challenge 3: Performance and Scalability}}

\begin{figure}[t]
	\centering
	\includegraphics[width=0.7\linewidth]{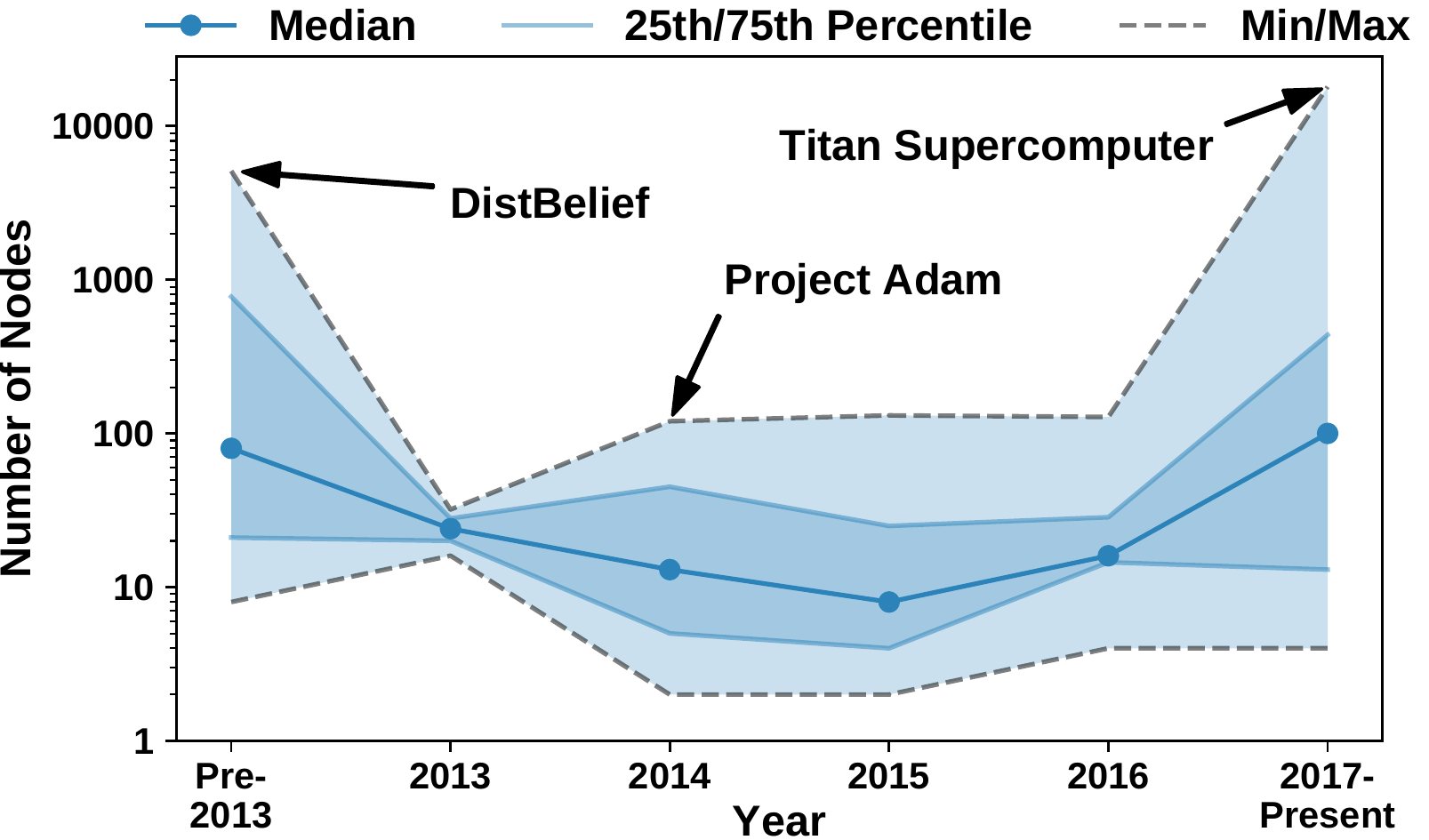}
	\caption{Statistics of using compute nodes in distributed DL~\cite{ben2018demystifying}}
	\label{fig:nodecount}
\end{figure}

Another unaddressed challenge is the design of distributed benchmarking of DL
to ensure high performance and scalability. As DL datasets and training
complexity continue to grow, large-scale distributed training is becoming an
essential DL component~\cite{ben2018demystifying} (Fig.~\ref{fig:nodecount}).
Every top competitor in DAWNBench~\cite{coleman2017dawnbench} uses multiple
multi-GPU nodes, and recently the entire Titan supercomputer (18,000 nodes) was
used for a full 24 hours to perform distributed DL via
meta-optimization (i.e., where the DNN structure may change)~\cite{young17titan}.  To deliver
high-quality scalable distributed DL codes, one must be able to debug
scalability issues, simultaneously preventing negative performance impact
coming from the benchmarking infrastructure.  Moreover, we need proper
techniques to understand such scalability bugs in the context of DL workloads.
As we show later (\cref{sec:evaluation}), \textbf{the Deep500 benchmarking infrastructure
potentially scales to thousands of cores and incurs negligible overheads
over native performance}.

\subsection{\textbf{Challenge 4: Validation}}

A benchmarking infrastructure for DL must allow to \emph{validate} results with
respect to several criteria.  As we discuss in~\cref{sec:evaluation}, \textbf{Deep500
offers validation of convergence, correctness, accuracy, and performance}.
Validation comes in the form of $\ell_1,\ell_2,\ell_\infty$ norms, but also in
forms of heatmaps, to highlight regions of interest, or repeatability via a map of output variance. In addition, we provide
gradient validation through numerical differentiation with similar metrics. We
also test optimizers in similar ways, making sure that optimization
trajectories do not diverge given the same inputs.

\subsection{\textbf{Challenge 5: Reproducibility}}

The final challenge in distributed DL benchmarking is the ability to
\emph{reproduce} or at least \emph{interpret}~\cite{hoefler2015scientific}
results. \textbf{In Deep500, we ensure these properties by using our
interfaces and several careful design decisions}, described
in~\cref{sec:design}.

\begin{figure*}[t]
\vspace{-0.5em}
\centering
\includegraphics[width=1.0\textwidth]{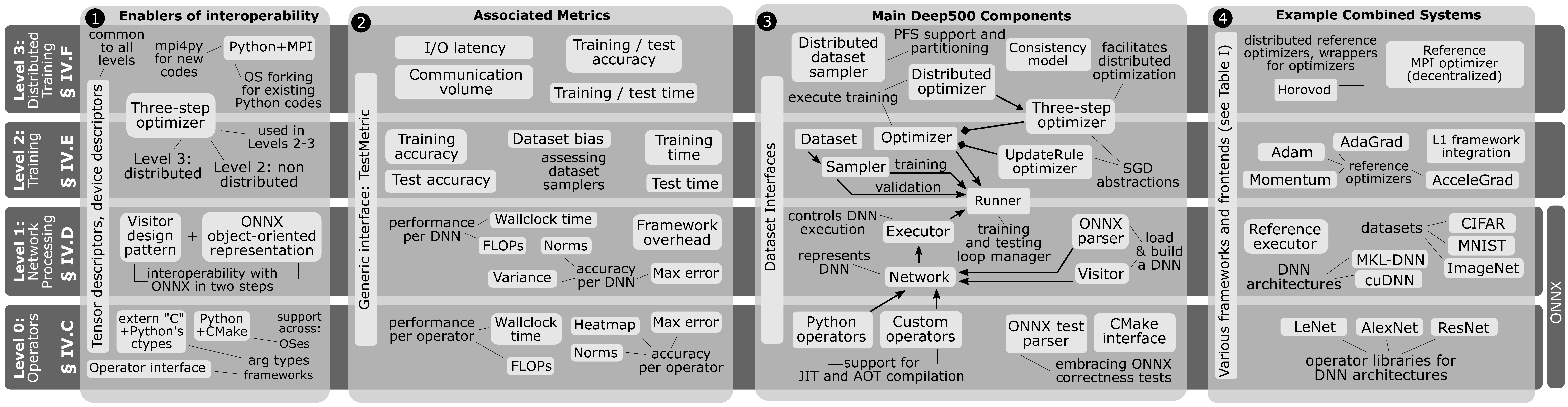}
\vspace{-1.5em}
\caption{The design of Deep500.}
\vspace{-1.5em}
\label{fig:design}
\end{figure*}

\section{Design and Implementation of Deep500}
\label{sec:design}
\label{sec:implementation}

We now describe the purpose and design of Deep500, addressing the above-discussed challenges.  \emph{The
core enabler} in Deep500 is \emph{the modular design} that groups all the
required functionalities into four \emph{levels}: \ding{182} ``Operators'', \ding{183}
``Network Processing'', \ding{184} ``Training'', and \ding{185} ``Distributed
Training''.  Each level provides relevant abstractions, interfaces, reference
implementations, validation procedures, and metrics.  We illustrate levels and
their relationships in Fig.~\ref{fig:dl}; the full design of the Deep500
meta-framework is shown in Fig.~\ref{fig:design}.

\subsection{Intended Purpose and Roles}

The Deep500 meta-framework is a benchmarking environment, and as such it is not meant to be a DL framework that provides optimized implementations of its own. Rather, Deep500 assumes high-performance frameworks exist. By abstracting the high-level aspects of DL (e.g., data loading) in a platform-agnostic manner, Deep500 enables the measurement and development of various metrics (performance, accuracy) in the different contexts of DL and distributed DL.

By taking the white-box approach, the user roles that Deep500 enables can be of a benchmark evaluator, or of an experimental scientist. In the former, one might use Deep500 and the various built-in metrics to choose hardware (or software) that performs best given a target workload. The latter role can use metrics and automatic integration with existing frameworks in order to empirically evaluate new operators, training algorithms, or communication schemes for DL. Since Deep500 provides reference code for nearly every concept, new methods can be validated against existing verified (yet slow) implementations. 

Deep500 complements existing DL frameworks in assisting
user efforts. DL frameworks and other integrations provide baselines for both performance and convergence, and all a user has to do is implement their own part, reusing the rest of the existing components. In the rest of this paper, we show that such extensions to the algorithms and metrics are simple to implement in a concise manner, and that the incurred overhead on performance is relatively low.

\subsection{Common Components}
\label{sec:components}

We first describe elements common to all four levels.

\macs{Metrics} 
Answering \myboxX{Use Cases 1, 3} and the resulting Challenge~2, Deep500
provides a general-purpose interface \emph{to access and use metrics}. In particular,
the \texttt{TestMetric} class contains methods to obtain the number of re-runs
needed for a selected measurement (e.g., used in ensuring numerical stability),
to make or summarize a measurement, and to generate a selected result (e.g., a
plot file or a series of numbers).

\macs{Interoperability: DNN Format}
To read and write DNNs, we use the ONNX format, and thus need to interoperate
with its Python package. As we show in more detail later (\cref{sec:level-1}), Deep500 
\emph{converts the ONNX format to an object-oriented notation for easier interoperability}. To support this, we
auto-generate code from the ONNX operator definition files.
We also extend ONNX with new operations for computing loss functions and optimization, as well as distributed optimization. Finally, we embrace the ONNX correctness tests.
This addresses \myboxX{Use Case~2} and Challenge~1: customizability and
interoperability with various data formats.
By using ONNX operators, we also address \myboxX{Use Case~1}:
Since ONNX standardizes a wide range of ML-related operators, Deep500 can be used to construct conforming DNNs between frameworks, such as TensorFlow and CNTK in the example.

\macs{Interoperability: Datasets and Networks}
Deep500 can download the MNIST~\cite{mnist}, Fashion-MNIST~\cite{fmnist}, and CIFAR-10/100~\cite{cifar}
datasets on demand, as well as parse ImageNet~\cite{imagenet}. 
Similarly, it facilitates access to DNN architectures (as ONNX files) for
LeNet~\cite{lenet}, ResNet~\cite{resnet} with varying depths, and Wide
ResNet~\cite{wideresnet}. 
Facilitating access to various datasets enhances DL programmability and
addresses \myboxX{Use Case~2}.

\macs{Interoperability: Frameworks and Platforms}
Deep500 uses its own \emph{descriptors} for \emph{tensors} and \emph{devices}
to enable interoperability with frameworks and platforms (\myboxX{Use Case~1}).  Tensor descriptors (\texttt{tensordesc}), which are also C
Application Binary Interface (ABI) compatible, extend the types given in ONNX
by describing the data type in more detail (e.g., allowing bitsets, or
including data layout types). These extensions enable each implemented
framework to convert types back and forth from Deep500 tensor descriptors.
Additionally, we provide extensible Device Descriptors (CPU, GPU, FPGA, etc.),
for example used to identify the most advantageous compute device for a
specific operator computation, addressing several
use cases from ``\myboxX{Others}''.

\macs{Interoperability: Distributed Training}
To facilitate distributed training and thus address \myboxX{Use Case~4} and
Challenge~3 (performance and scalability), we use
a Python interface with MPI, \texttt{mpi4py}, \emph{to link with MPI}, and use
OS forking to turn an existing Python application into an MPI-capable one, all
while keeping the proper distributed DL semantics w.r.t. dataset sampling and
distributed storage, as well as to the DNN model.

\subsection{Level 0: Operators}
\label{sec:level-0}

Level~0 enables implementing, computing, and benchmarking individual operators,
which are the building blocks of DNNs. An operator's functionality is general, and spans DNN layers (e.g., convolution) as well as training-related operations (e.g., distributed gradient accumulation).

\macs{Interfaces} 
Deep500 Level~0 allows to integrate new custom operators
with real datasets, networks, or frameworks, without having to implement other
operators. For this, we provide the \texttt{CustomOperator} interface,
available in Python (addressing high-level ML researchers and experimentation)
or in C++ (addressing high-performance implementations).
\texttt{CustomOperator} provides two functions,
\texttt{forward(inputs)} and \texttt{backward(grad\_inputs, fwd\_inputs,
fwd\_outputs)}. To support the integration of arbitrary C++ code in existing frameworks and abstractions, we provide a runtime \textit{compilation interface}. The compiler interface is a simple wrapper around CMake, which includes stub templates for each implemented DL framework. These templates include the custom C++ code to create a framework-compatible interface of an operator, which can be seamlessly used in the frameworks, even without the rest of Deep500.
Using this abstraction, Deep500 supports both Just-In-Time (JIT) or Ahead-Of-Time (AOT) compilation of operators, enabling flexible benchmarking of high-performance code.

\macs{Example Use Case} 
Listings~\ref{lst:example-level-0-c++}--\ref{lst:example-level-0-python}
illustrate Deep500's interoperability with frameworks: they
contain an example definition of a custom \emph{median-pooling} operator in C++
and its straightforward registration as well as compilation for PyTorch.

\begin{lstlisting}[language=c++, float=h, label=lst:example-level-0-c++,
caption=(\cref{sec:level-0}) Defining a custom operator in Deep500 with C++.]
|\tikzmarkin{col-0-0}(8.5,-0.175)(-0.0,0.25)|template<typename T>
class MedianPooling : public deep500::CustomOperator {
public:
  void forward(const T *input, T *output) { /* Inference code */ }
  void backward(const T *nextop_grad, const T *fwd_input_tensor,
                const T *fwd_output_tensor, T *input_tensor_grad) {
    /* Backpropagation code */ }
};|\tikzmarkend{col-0-0}|                                                |\myboxX{\textbf{ Operator definition }}|

|\tikzmarkin{col-0-1}(5,-0.175)(-0.0,0.27)|D500_REGISTER_OP(MedianPooling<DTYPE>); //Register a custom operator
 D500_EXPORTED void *create_new_op(deep500::tensor_t *input_descs,
    int ninputs, deep500::tensor_t *output_descs, int noutputs) {
  //Create the actual operator object.
  return new MedianPooling<DTYPE>(/* ... */);
}                             |\tikzmarkend{col-0-1}|                   |\myboxX{\textbf{ Operator registration }}|
\end{lstlisting}

\begin{lstlisting}[language=python, float=h, label=lst:example-level-0-python,
caption=(\cref{sec:level-0}) Using a custom operator in Deep500 with Python.]
import deep500 as d5
from deep500.frameworks import pytorch as d5pt|\vspace{0.5em}|
|\tikzmarkin{col-1p-0}(8.8,-0.175)(-0.0,0.25)|# Create an operator descriptor for compilation  |\myboxX{\textbf{ Operator compilation }}||\vspace{0.5em}|
 opdesc = d5.compile_custom_cppop('MedianPooling', 'mp.cpp', 
    [d5.tensordesc(tf.float32, [256, 256])], # Input tensor shapes
    [d5.tensordesc(tf.float32, [128, 128])], # Output tensor shapes
    additional_definitions={'DTYPE': 'float'})
|\tikzmarkend{col-1p-0}| mycppop = d5pt.custom_op(opdesc) # Compiles operator for framework
\end{lstlisting}

\macs{Interoperability}
When an operator has more than one input or output, supporting a
high-performance C++ implementation is complicated.  Arrays incur overheads due
to dynamic memory management and the code becomes less readable. To solve this
issue, Deep500 uses \emph{variable} arguments in the C++ interface, then
exports a C \emph{ABI-compatible} function (\texttt{extern "C"}, containing no
defined arguments), and uses Python's dynamic library invocation interface
(\texttt{ctypes}) to call the function \emph{with unpacked arguments}. We also
automatically convert native tensor types into C pointers, so that \emph{any}
operator is implemented \emph{only once} for \emph{all} frameworks.

To support custom C++ operators across frameworks \emph{and} OSes
(Deep500 supports Linux, Windows, and Mac OS), we use CMake. To enable JIT/AOT
compilation of operations, we wrap CMake process with a cross-platform
Python interface that can accept multiple files for compilation.

Finally, our operator interface allows to convert native operators from
frameworks into custom Deep500 operators for use in, e.g., other frameworks,
see Listing~\ref{lst:level-0-native-custom}.

\begin{lstlisting}[language=python, float=h, label=lst:level-0-native-custom,
caption=(\cref{sec:level-0}) Using a native operator as a custom one in Deep500 (Python).]
import tensorflow as tf
from deep500.frameworks import tensorflow as d5tf|\vspace{0.5em}|
|\tikzmarkin{col-1c-0}(8.8,-0.175)(-0.0,0.25)|# ...Define A, B and C as TensorFlow tensors   |\myboxX{\textbf{ Operator customization }}||\vspace{0.5em}|
op = d5tf.custom_op_from_native(tf.matmul, 
    [d5tf.desc_from_tensor(A), d5tf.desc_from_tensor(B)],
    [d5tf.desc_from_tensor(C)])
|\tikzmarkend{col-1c-0}|# Deep500 can now use the operator interface to test `op`
\end{lstlisting}

\macs{Provided Implementations}
Deep500 provides reference implementations of all operators required for the DNNs
in~\cref{sec:components}.

\macs{Metrics}
One family of the associated metrics are performance per operator, for example
wallclock time, FLOPs, or consumed energy. Another family are accuracy per
operator: norms (e.g., $\ell_1,\ell_2,\ell_\infty$), variance in output, and
2D/3D heatmaps.

\macs{Validation}
Validation is enabled by two functions.
First, \texttt{test\_forward} tests operator correctness
and performance.  Second, \texttt{test\_gradient} uses numerical differentiation (Jacobian matrix evaluation using
finite differences) to provide automatic gradient checking, as well as measure
the performance of the \texttt{backward} method.

\begin{figure}[t]
	\centering
	\includegraphics[width=1.0\columnwidth]{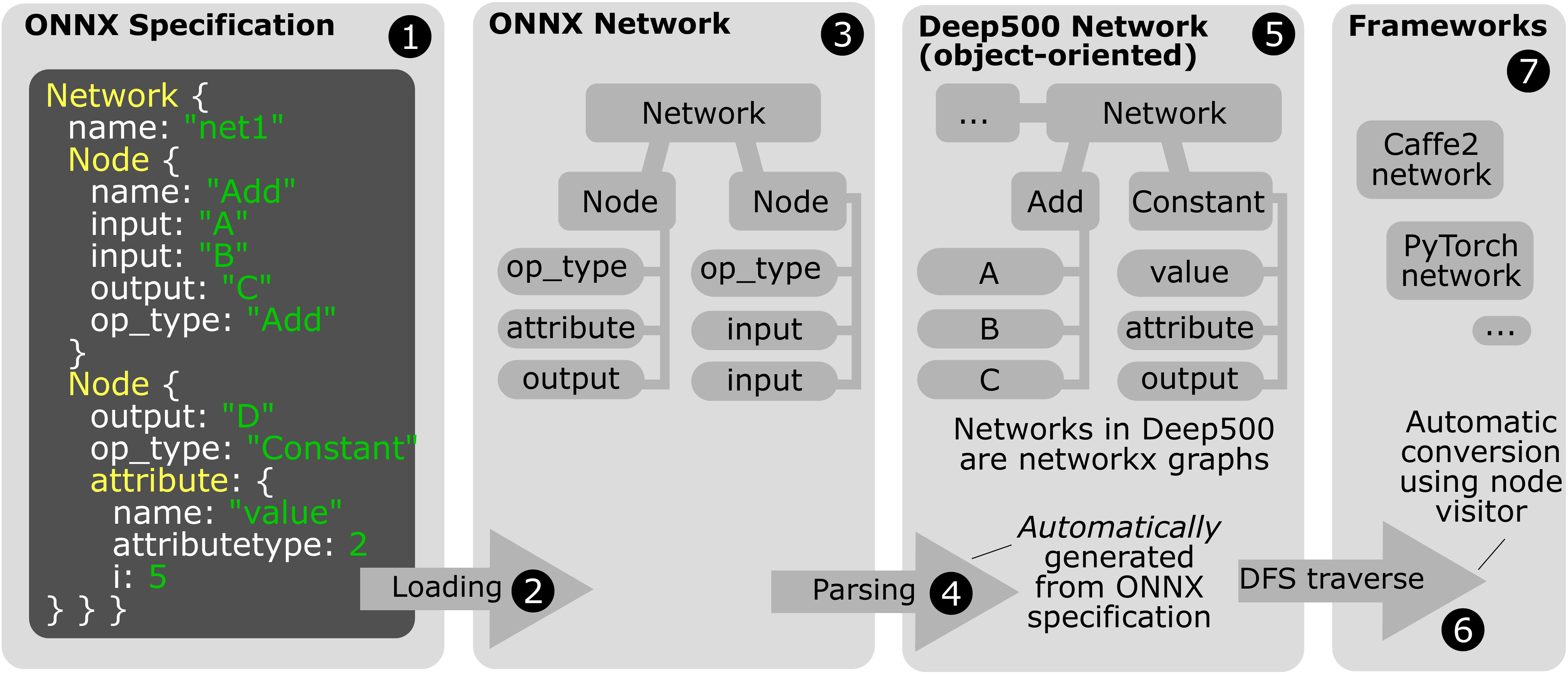}
	\vspace{-1em}
	\caption{An automatic transformation of an ONNX network
		to a Deep500 network.}
	\label{fig:onnxvisitor}
\end{figure}

\subsection{Level 1: Network Processing}
\label{sec:level-1}

Level 1 is dedicated to the construction, modification, evaluation, and
backpropagation of entire neural networks.
Deep500 separates the network abstraction from file formats, operators,
and training to enable a fair and extensible infrastructure upon which
researchers can build their own graph transformations to optimize between
operators.

\macs{Interfaces}
To represent a DNN, we use two classes: \texttt{Network} and
\texttt{GraphExecutor}. \texttt{Network} defines the network structure and
exposes a standard graph API that allows to \texttt{add} and \texttt{remove}
nodes/edges, \texttt{fetch} node data contents,
\texttt{feed} nodes with new values, and others. 
\texttt{GraphExecutor} controls the DNN execution. It provides two functions
that enable inference and possibly backpropagation: \texttt{inference}, and
\texttt{inference\_and\_backprop}. 

To enable fine-grained measurements and support early exits, graph executors
must be able to invoke certain \texttt{Event}s at the right time. Events are
user-specified ``hooks'' that are called at certain points during complex
actions such as backpropagation and training.  An example of an event ``hook''
is providing an early stopping condition.  To enable benchmarking events with a
selected metric, the same metric class can extend both the \texttt{TestMetric}
and \texttt{Event} classes.

\macs{Interoperability}
While \texttt{Network}s can be created manually (node by node), Deep500
also provides a convenient interface to construct networks from ONNX
files. This entails a non-trivial processing scheme, depicted in
Fig.~\ref{fig:onnxvisitor}, in which an ONNX graph is first transformed to an
intermediate, object-oriented representation. Deep500 then uses the
\emph{Visitor design pattern} to invoke \texttt{Network} construction by
calling the right functions. An example construction is in
Listing~\ref{lst:level-1-visitor}.

\begin{lstlisting}[language=python, float=h, label=lst:level-1-visitor, caption=An example DNN construction using the TensorFlow ONNX visitor.]
class TensorflowVisitor(d5.OnnxBaseVisitor):
  # ... other definitions ... 
|\tikzmarkin{col-1-0}(8.8,-0.175)(-0.0,0.23)| def visit_dropout(self, op: d5.ops.Dropout, network: TFNetwork):
    X = network.fetch_internal_tensor(op.i_data)
    ratio = op.ratio.get_value() if op.ratio else 0.5
    Y = tf.nn.dropout(X, ratio)
|\tikzmarkend{col-1-0}|    network.feed_internal_tensor(op.o_output, Y)      |\myboxX{\textbf{ Dropout visitor }}| 

|\tikzmarkin{col-1-1}(8.8,-0.175)(-0.0,0.23)|  def visit_sub(self, op: d5.ops.Sub, network: TFNetwork):
    A, B = network.fetch_internal_tensors([op.i_A, op.i_B])
    C = tf.subtract(A, B)
|\tikzmarkend{col-1-1}|    network.feed_internal_tensor(op.o_C, C)       |\myboxX{\textbf{ Subtraction visitor }}| 

|\tikzmarkin{col-1-2}(8.8,-0.15)(-0.0,0.23)|  def visit_mul(self, op: d5.ops.Mul, network: TFNetwork):
    A, B = network.fetch_internal_tensors([op.i_A, op.i_B])
    C = tf.multiply(A, B)
|\tikzmarkend{col-1-2}|    network.feed_internal_tensor(op.o_C, C)     |\myboxX{\textbf{ Multiplication visitor }}| 
\end{lstlisting}

\macs{Provided Implementations}
Deep500 implements a reference \texttt{Network} using the \texttt{networkx} Python graph
library. \texttt{GraphExecutor} is implemented by a topological graph sort.

\macs{Metrics}
We adapt the metrics from Level~0 to full DNN execution. We also add
the \texttt{FrameworkOverhead} metric, which measures the overall time for
inference and backpropagation and compares it with the sum of running times of
individual operators. This evaluates the impact from framework and hardware
management (e.g., GPU kernel invocation latency).

\macs{Validation}
To validate the accuracy and performance of \texttt{Network} and
\texttt{GraphExecutor}, we provide two functions:
\texttt{test\_executor}, and
\texttt{test\_executor\_backprop} for inference and
backpropagation, respectively.

\begin{figure*}[t]
\vspace{-0.5em}
	\centering
	\includegraphics[width=1.0\textwidth]{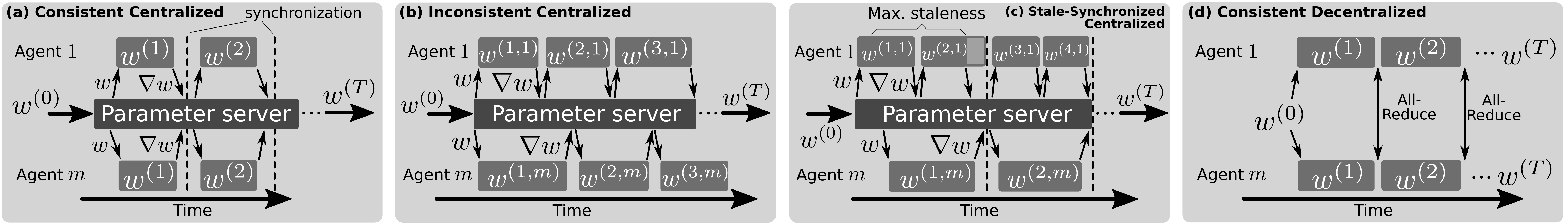}
	\vspace{-1.5em}
   \caption{Examples of distributed deep learning schemes ($w^{(x,m)}$ are
   minimizing parameters in iteration $x$ belonging to agent~$m$;
   see~\cref{sec:dl} for more details).} \label{fig:distdl}
	\vspace{-1.5em}
\end{figure*}

\subsection{Level 2: Training}

Level~2 of Deep500 implements DNN training.

\macs{Interfaces} 
The main Level~2 interfaces are \texttt{DatasetSampler} and \texttt{Optimizer}.
First, \texttt{DatasetSampler} provides minibatches by sampling a given
dataset, and can be extended to test different sampling schemes. 
For more performance, samplers can be implemented as custom operators in C++
and plugged into a DNN \texttt{Network} as native operators. Existing native
sampler operators, such as \texttt{tf.Dataset} in TensorFlow, can also
seamlessly be used.
The second main interface, \texttt{Optimizer}, uses a \texttt{GraphExecutor}
and a \texttt{DatasetSampler}, and can run any code as the training procedure.
We provide two abstractions of SGD optimizers: \texttt{UpdateRuleOptimizer},
which runs an update rule akin to $U$ in Algorithm~\ref{alg:sgd}, and
\texttt{ThreeStepOptimizer}, a novel abstraction that
facilitates \emph{distributed} optimization. To facilitate automatic
distribution of optimization, we divide \texttt{optimizer}'s execution into
three steps: \ding{182} input sampling (Algorithm~\ref{alg:sgd},
line~\ref{alg:sgd:readsamples}), \ding{183} adjusting parameters prior to
inference (line~\ref{alg:sgd:readweights}), and \ding{184} applying an update
rule (line~\ref{alg:sgd:accgrad}).

\macs{Interoperability} 
Implementing new optimizers in DL frameworks, especially ones that do not
conform to simple update rules, is a notoriously hard task. This hinders
testing new algorithms with good convergence guarantees~\cite{levy2018online}
or non-SGD methods such as second-order optimization.  Additionally,
interfacing with datasets is not standardized across frameworks, with varying
data augmentation methods, and minibatch sampling being hardcoded into
frameworks. 
Deep500 alleviates these issues with the \texttt{ThreeStepOptimizer} interface.

\macs{Example Use Case} 
Listing~\ref{lst:level-1-accelegrad} illustrates the \textit{full}
implementation of AcceleGrad, a state-of-the-art DL optimizer~\cite{levy2018online}, using 
Deep500's \texttt{ThreeStepOptimizer}. It is apparent that the code retains its algorithmic form.

\begin{lstlisting}[language=python, float=h, label=lst:level-1-accelegrad, caption=
Implementation of AcceleGrad in Deep500.]
class AcceleGradOptimizer(d5.ThreeStepOptimizer):|\vspace{0.5em}|
|\tikzmarkin{col-2-0}(8.8,-0.175)(-0.0,0.25)| def new_input(self):                                     |\myboxX{\textbf{ New input }}||\vspace{0.5em}|
    self.t = self.t + 1
    self.alpha_t = 1 if 0 <= self.t <= 2 else 1 / 4 * (self.t + 1)
|\tikzmarkend{col-2-0}|    self.tau_t = 1 / self.alpha_t

|\tikzmarkin{col-2-1}(8.8,-0.175)(-0.0,0.25)| def prepare_param(self, param_name):              |\myboxX{\textbf{ Adjust parameters }}||\vspace{0.5em}|
    param = self.executor.network.fetch_tensors([param_name])[0]
    if not self.init:
      self.y[param_name] = param
      self.z[param_name] = param
      self.squares[param_name] = 0
    y = self.y[param_name]
    z = self.z[param_name]
    new_param = self.tau_t * z + (1 - self.tau_t) * y
|\tikzmarkend{col-2-1}|    self.executor.network.feed_tensor(param_name, new_param)

|\tikzmarkin{col-2-2}(8.8,-0.175)(-0.0,0.25)| def update_rule(self, grad, old_param, param_name):
    squared_grad = self.squares[param_name]
    squared_grad += self.alpha_t ** 2 * np.linalg.norm(grad) ** 2
    eta_t = 2 * self.D / np.sqrt(self.G ** 2 + squared_grad)
    z_t = self.z[param_name]
    z_t2 = z_t - self.alpha_t * eta_t * grad
    y_t2 = old_param - eta_t * grad
    self.z[param_name] = z_t2
    self.y[param_name] = y_t2
    self.squares[param_name] = squared_grad
    adjusted_lr = self.lr / (self.eps + np.sqrt(squared_grad))
    self.init = False
    
|\tikzmarkend{col-2-2}|    return old_param - adjusted_lr * grad                |\myboxX{\textbf{ Update rule }}| 
\end{lstlisting}

\macs{Provided Implementations}
Deep500 provides many popular optimizers written in Python, such as Gradient Descent with
learning rate schedule, momentum, Adam~\cite{adam}, and AdaGrad.

\macs{Metrics}
Deep500 provides two main metrics in Level~2: \texttt{TrainingAccuracy}
(measures the training accuracy at every $k$th step) and
\texttt{TestAccuracy} (measures the test accuracy at every $k$th epoch).
Additionally, dataset samplers can be tested \emph{individually} by running
\texttt{test\_sampler} with the \texttt{DatasetBias} metric, which collects a
histogram of sampled elements w.r.t. corresponding labels.

\macs{Validation}
First, \texttt{test\_optimizer} verifies the performance and correctness of
\emph{one step} of the optimizer (ensuring that an optimizer trajectory does
not diverge from the Deep500 one). Second, \texttt{test\_training} tests the
convergence, performance, and the related tradeoff of the \emph{overall}
training.

\subsection{Level 3: Distributed Training}

A \emph{cornerstone feature} of Deep500 is that \emph{it distributes} 
DNN training \emph{with virtually no effort from an API user}.
The core enablers are the two class interfaces from Level~2: an update-rule
optimizer and a three-step optimizer. The distributed MPI-based optimizer uses these
classes to distribute new gradients and parameters before or after
executing update rules.

\macs{Interfaces} 
We provide two interfaces: \texttt{DistributedSampler} and
\texttt{DistributedOptimizer}. The former refers to a potentially distributed
data store.  The latter refers to a specific subclass of \texttt{Optimizer}s
from Level~2, which support distributed communication.  Still, there are no
pre-defined constraints on \textit{how} and \textit{when} communication should
occur, so this could potentially be a gradient-free optimizer, e.g., employing
Genetic Algorithms~\cite{regevo,young17titan}. The three-step and update-rule
optimizers from Level~2 also extend \texttt{DistributedOptimizer}, so
implementing a custom optimizer based on these methods \emph{automatically
grants distribution capabilities}.

\macs{Example Use Case}
Testing for cluster-wide performance of different communication and parameter consistency schemes are notoriously hard tasks. In most codes, they require an almost total re-write of the program logic, depend on additional libraries (sometimes entailing \textit{framework recompilation}), and not supported on all platforms, especially supercomputers. Due to the modularity of Deep500, the task is a matter of wrapping an optimizer with the right distributed scheme (Listing~\ref{lst:level-3-cluster-wide}).

\begin{lstlisting}[language=python, float=h, label=lst:level-3-cluster-wide,
caption=Testing cluster-wide performance of distributed training in Deep500.]
gexec = # ... (the definition uses previous levels)
opt = d5ref.AdamOptimizer(gexec)
ds = d5ref.ShuffleDistributedSampler(dataset, batch_size)
# Collect metrics for different training schemes and topologies
ref = d5.test_training(d5ref.ConsistentDecentralized(opt), ds)
ps = d5.test_training(d5ref.ConsistentCentralized(opt), ds)
asgd = d5.test_training(d5ref.InconsistentCentralized(opt), ds)
hvd = d5.test_training(d5tf.HorovodDistributedOptimizer(opt), ds)
\end{lstlisting}

\macs{Interoperability}
Deep500 also facilitates the construction of new communication methods,
topologies, and interfaces. None of these features are natively supported by
frameworks.  For example, modifying a DNN graph to create pipeline parallelism
across processes is impossible automatically in \emph{any} of the frameworks,
but can straightforwardly be done in Deep500.

\macs{Provided Implementations}
Deep500 implements different distributed SGD variants.
This includes centralized and decentralized optimization
(\cref{sec:dl}), a variant with a globally-consistent model (synchronous
SGD)~\cite{ben2018demystifying}, inconsistent model (asynchronous SGD, e.g.,
HOGWILD~\cite{recht2011hogwild}), and stale-synchronous SGD, which enables inconsistency
in the parameters up to a certain delay. Fig.~\ref{fig:distdl} illustrates
possible timelines of each method. These methods are, of course, compatible
with all frameworks, as they use the MPI library separately. 
Listing~\ref{lst:level-3-opt} illustrates
the achieved compatibility using the \textit{full} implementation of the
Consistent Decentralized optimizer.

\begin{figure*}[t]
%\vspace{-1em}
\centering
	\begin{subfigure}{.46\textwidth}
		\includegraphics[trim={0.55cm 0.55cm 0.55cm 0.5cm},clip,width=0.49\textwidth]{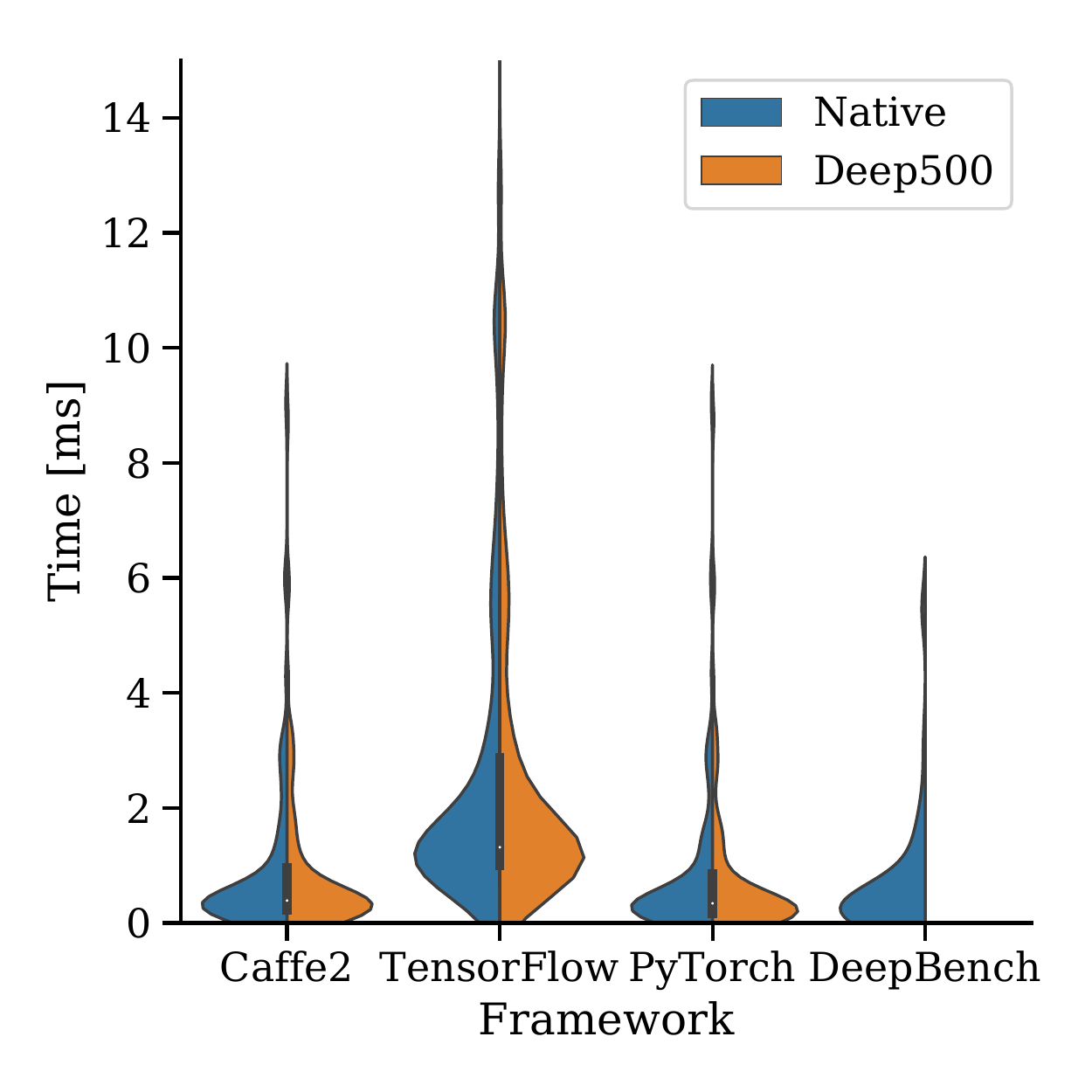}
		\includegraphics[trim={0.55cm 0.55cm 0.55cm 0.5cm},clip,width=0.49\textwidth]{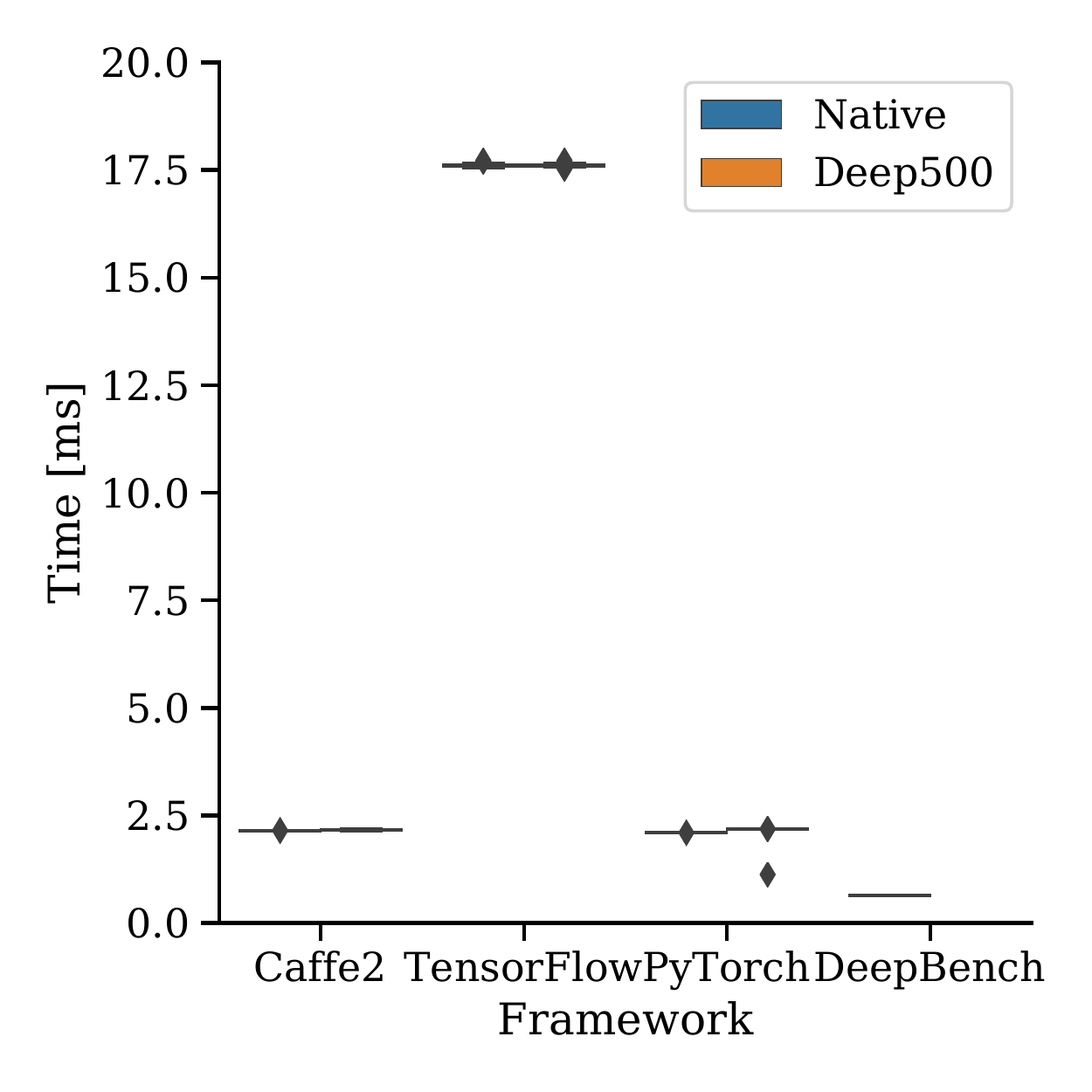}
		\vspace{-1em}
		\caption{Convolution Performance (left: violin plot of all kernels, right: box plot of size $N=16$, $C=3$, $H=W=224$, filter size $3\times 3$).}
		\label{fig:micro-timing-error-conv}
	\end{subfigure}
	\qquad
	\begin{subfigure}{.46\textwidth}
		\includegraphics[trim={0.55cm 0.55cm 0.55cm 0.5cm},clip,width=0.49\textwidth]{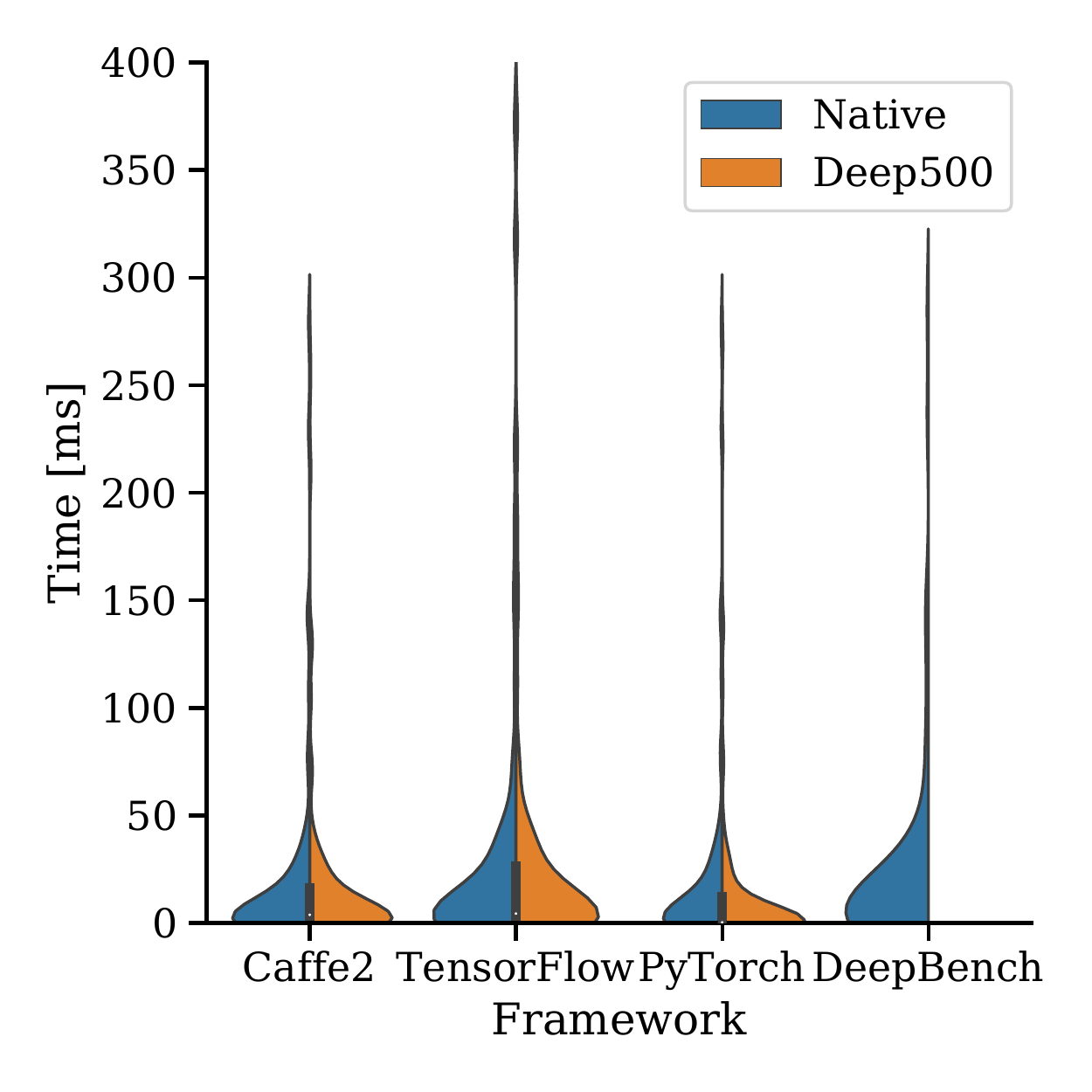}
		\includegraphics[trim={0.55cm 0.55cm 0.55cm 0.5cm},clip,width=0.49\textwidth]{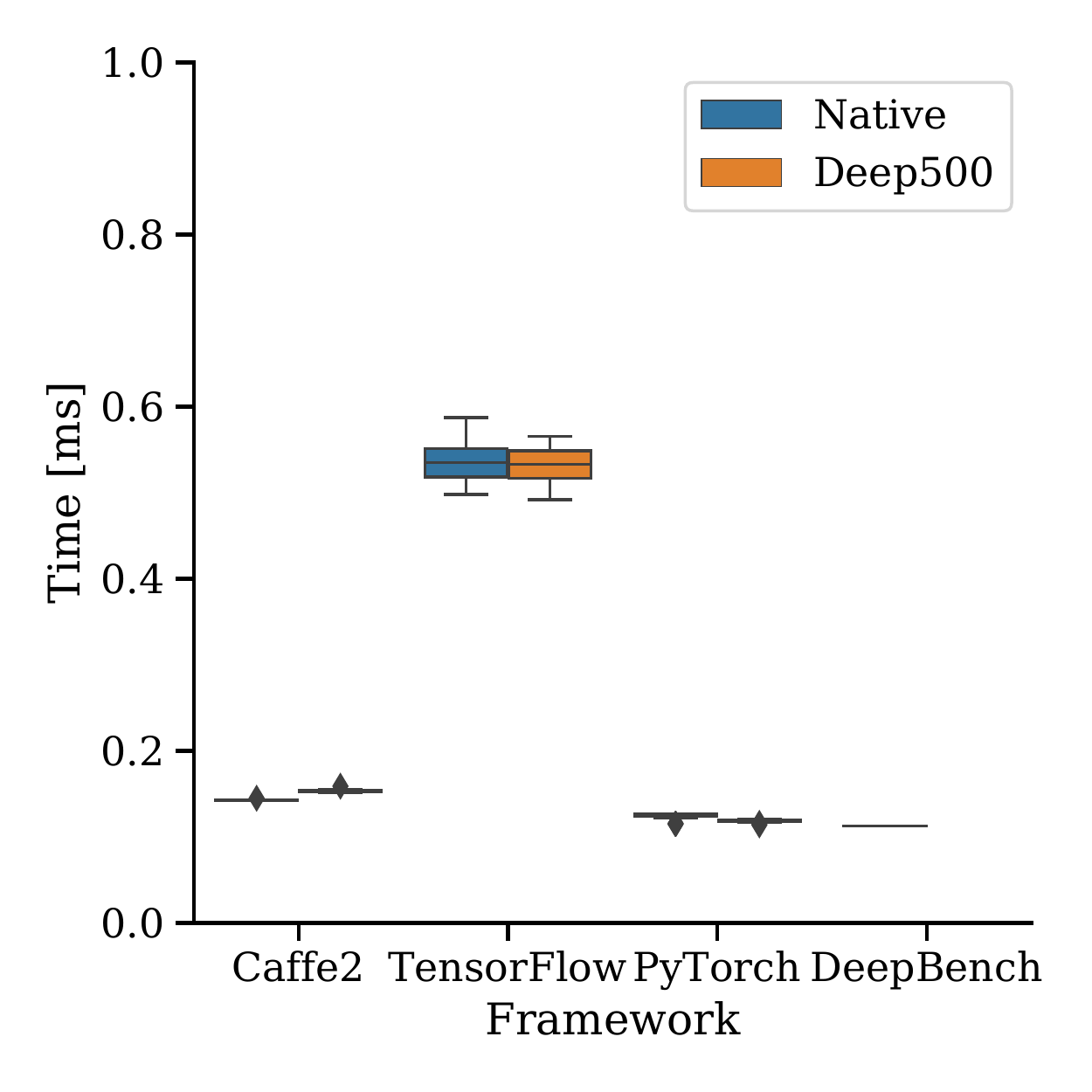}
		\vspace{-1em}
		\caption{Matrix Multiplication Performance (left: violin plot of all kernels, right: box plot of size $M=K=2560$, $N=64$).}
		\label{fig:micro-timing-error-gemm}		
	\end{subfigure}
\vspace{-0.25em}
\caption{\textbf{Analysis of Deep500 Level~0:} Performance of operators
implemented with Deep500 and selected frameworks, together with
the DeepBench baseline.}
\label{fig:micro-timing-error}
\end{figure*}

\begin{lstlisting}[language=python, label=lst:level-3-opt, float=h,
caption=Consistent Decentralized reference optimizer in Deep500.]
class ConsistentDecentralized(DistributedOptimizer):
  # self.base_optimizer is a `ThreeStepOptimizer` object
  def train(self, inputs):
    self.base_optimizer.new_input()
    for param in self.network.get_params():
      self.base_optimizer.prepare_param(param) 
    output = self.executor.inference_and_backprop(inputs)
    gradients = self.network.gradient()
    for pname, grad_name in gradients:
      param, grad = self.network.fetch_tensors([pname, grad_name])
      grad = self.communication.sync_all(grad)
      param = self.base_optimizer.update_rule(grad, param, pname)
      self.network.feed_tensor(pname, param)
    return output
\end{lstlisting}

As opposed to native distributed optimization, using reference implementations
in conjunction with a GPU incurs a synchronous GPU-to-host copy prior to
communicating, and vice versa. This could be further improved by custom C++
operators that implement a specialized \texttt{forward\_cuda} method, e.g.,
using CUDA-aware MPI or GPUDirect.

\macs{Metrics}
We provide two metrics: \texttt{CommunicationVolume} and \texttt{DatasetLatency}.

\macs{Validation}
We reuse the two validation functions from Level~2: \texttt{test\_optimizer}
and \texttt{test\_training}. However, instead of an \texttt{Optimizer}
and a \texttt{DatasetSampler}, we feed \texttt{DistributedOptimizer} and
\texttt{DistributedSampler} classes.

\section{Evaluation}
\label{sec:evaluation}

Our key goal in this section is to show that \textbf{Deep500 enables detailed,
accurate, and customizable benchmarking of DL codes while incurring negligible
overheads}. 

\subsection{Methodology, Setup, Parameters}

\macs{Neural Networks}
Operator dimensions and types for Level~0 tests were collected from the
DeepBench~\cite{deepbench} low-level benchmark. For convergence tests, we
use ResNet-18 and 50~\cite{resnet}.
We use the small datasets MNIST~\cite{mnist} and CIFAR-10~\cite{cifar}, as well as the large-scale Imagenet~\cite{imagenet} dataset, where the latter uses JPEG files packed in the \texttt{TFRecord} file format.

\macs{Experimental Setup and Architectures}
We use the {CSCS Piz Daint} supercomputer. Each XC50 compute node contains a
12-core HT-enabled Intel Xeon E5-2690 CPU with 64 GiB RAM, and one NVIDIA Tesla
P100 GPU. The nodes communicate using Cray's Aries interconnect.

\macs{Evaluation Methodology}
To gather data for the non-distributed experiments (Levels 0--2), we run them
30 times and report median results and nonparametric 95\% confidence intervals.
We use 32-bit (single precision) floating point values for all DNN parameters
and errors. 

In all following benchmarks, Deep500 incurs certain overheads caused
by additional data copying while conducting measurements and recording the outcomes.
We expect that --- as with any other benchmarking infrastructure --- Deep500 users
would switch off unnecessary benchmarking metrics and instrumentation for production runs and other performance-critical scenarios.

\subsection{Level 0: Operators}

We first investigate \emph{performance and accuracy of
operators} implemented with Deep500, PyTorch, TensorFlow, and Caffe2. We also
consider NVIDIA \emph{native results} obtained with DeepBench. DeepBench
provides 160 different matrix multiplication sizes and 94 convolution
dimensions, typically found in DL workloads. We aggregate the results and present the
running time distribution and accuracy of each framework.

We seek to show that \textbf{Deep500's Level~0 is reliable and fast.}
Indeed, Fig.~\ref{fig:micro-timing-error} shows that for all frameworks,
\emph{Deep500 operator runtime differs negligibly (within CIs) from the native frameworks}. 
Moreover, DeepBench can be used as a baseline for operator runtime (assuming all frameworks are implemented over the same low-level libraries, such as cuDNN), as it only calls
a given kernel and outputs the
resulting GPU runtime. This is not the case in other frameworks, as they contain management overhead and additional actions, such as copying tensors.

Since convolution and matrix multiplications vary widely in size and runtime, we choose two common problem sizes from DeepBench and present the results in the right-hand side of Figures \ref{fig:micro-timing-error-conv} and \ref{fig:micro-timing-error-gemm}. The figures show that DeepBench is indeed faster than all frameworks, and the processing time varies between frameworks, where TensorFlow is the slowest and PyTorch is on average the fastest. The trends in a single example are similar to the overall results, however, TensorFlow and PyTorch over Deep500 are slightly faster than their native counterparts. Upon closer inspection, though, the runtime distributions are statistically indistinguishable.

As for correctness, the median (over the set of problem sizes) of the $\ell_\infty$ norms between Deep500 and
the frameworks are $\approx$0.0007, $\approx$0.00068, and $\approx$0.00073 for
TensorFlow, Caffe2, and PyTorch respectively.

\subsection{Level 1: Networks}

\begin{figure}[h!]
	\vspace{-1em}
	\centering
	\includegraphics[width=1.0\linewidth]{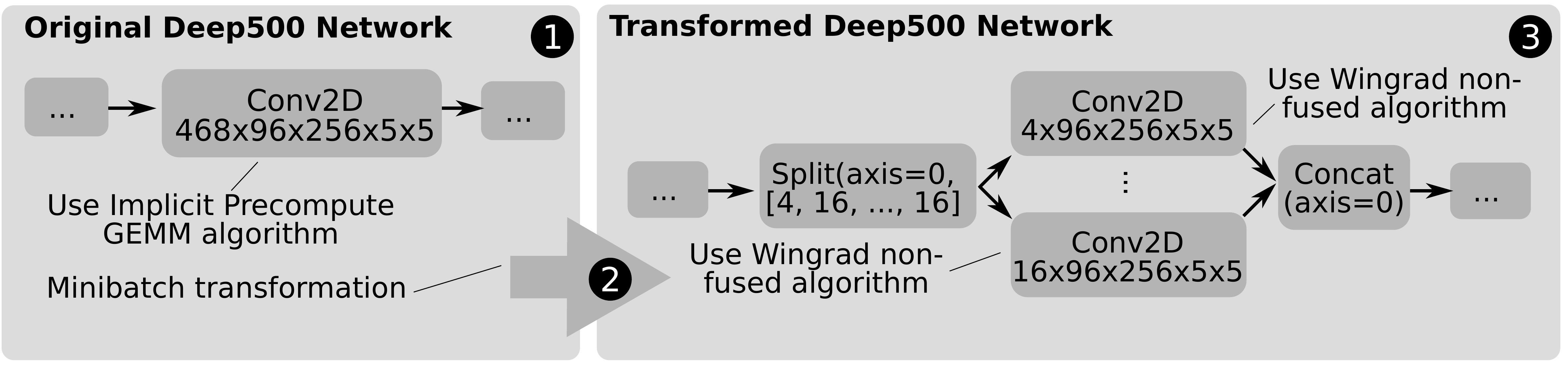}
	\vspace{-1.5em}
	\caption{The illustration of the Microbatch transformation.}
	\label{fig:microbatch}
	\vspace{-0.75em}
\end{figure}

In the Level~1 analysis, we investigate the performance and accuracy benefits
of \emph{transforming a DNN convolution} by splitting input minibatches into
smaller micro-batches, as Oyama et al.
propose~\cite{microbatch}. 
We apply the transformation on the network \emph{independently of
the framework} by solving an Integer Linear Program (ILP) to maximize
performance and preserve memory utilization constraints. The Level~1 code then replaces
convolutions with a split, followed by
micro-batch convolution and concatenation operations, as illustrated in
Fig.~\ref{fig:microbatch}.

We use Deep500 to apply this transformation on both PyTorch and TensorFlow.
Before the transformation, PyTorch suffered out-of-memory (OOM) errors for
AlexNet~\cite{alexnet} for minibatch sizes of 468 or higher. Deep500's
transformation decreases memory requirements, \textbf{eliminating OOM issues
and enabling PyTorch to process a given dataset} in $\approx$225ms.
Yet, the same optimization slows down TensorFlow from $\approx$350ms to
$\approx$380ms, because splitting and concatenating nodes in TensorFlow incur
additional memory copies. To alleviate this, one could modify the ONNX format
to allow \emph{referencing} sub-tensors instead of copying them.

%        \begin{table}[h]
%        %\scriptsize
%        \ssmall
%        \sf
%        \renewcommand{\arraystretch}{0.8}
%        \centering
%        %
%        %\resizebox{\textwidth}{!}{%
%        %
%        \begin{tabular}{lll}
%        \toprule
%        \textbf{Epoch} & \multicolumn{1}{c}{\makecell[l]{\textbf{TensorFlow}\\\textbf{(native)}}} & \multicolumn{1}{c}{\makecell[l]{\textbf{TensorFlow}\\\textbf{(Deep500)}}} \\
%        \midrule
%        %
%        % 0 & 247.2994971275 & 254.2741508484 \\
%        % 1 & 242.9404921532 & 242.9453964233 \\
%        % 2 & 242.8943395615 & 242.9795174599 \\
%        % 3 & 242.9469845295 & 242.9351446629 \\
%        % 4 & 242.9235265255 & 242.9188194275 \\
%        % 5 & 243.021951437  & 242.9739191532 \\
%        % 6 & 242.9277544022 & 242.9587442875 \\
%        % 7 & 242.9519767761 & 242.9454965591 \\
%        % 8 & 243.0073399544 & 242.92359519   \\
%        % 9 & 242.9535527229 & 243.0242242813 \\
%        % %
%        0 & 247.299 & 254.274 \\
%        1 & 242.940 & 242.945 \\
%        2 & 242.894 & 242.980 \\
%        3 & 242.947 & 242.935 \\
%        4 & 242.924 & 242.919 \\
%        5 & 243.022  & 242.974 \\
%        6 & 242.928 & 242.959 \\
%        7 & 242.952 & 242.945 \\
%        8 & 243.007 & 242.924   \\
%        9 & 242.954 & 243.024 \\
%        %
%        \bottomrule
%        \end{tabular}
%        %
%        %}
%        %
%        \caption{Native TensorFlow vs. TensorFlow integrated with Deep500 [ms].}
%        \label{tbl:native_vs_deep500}
%        \end{table}

\subsection{Level 2: Training}

In the Level~2 analysis, \emph{we compare the performance and accuracy of
training using TensorFlow, Caffe2, and Deep500 reference optimizers}.

\macs{Optimization Overhead}
We first measure the runtime of training in native TensorFlow and using the Deep500 TensorFlow integration. 
Apart from an instantiation overhead in the first epoch, Deep500 consistently incurs negligible ($<$1\%)
overhead, where both implementations take $\approx$243ms per epoch.

\begin{figure}[h!]
	\vspace{-1em}
	\centering
	\includegraphics[width=0.42\linewidth]{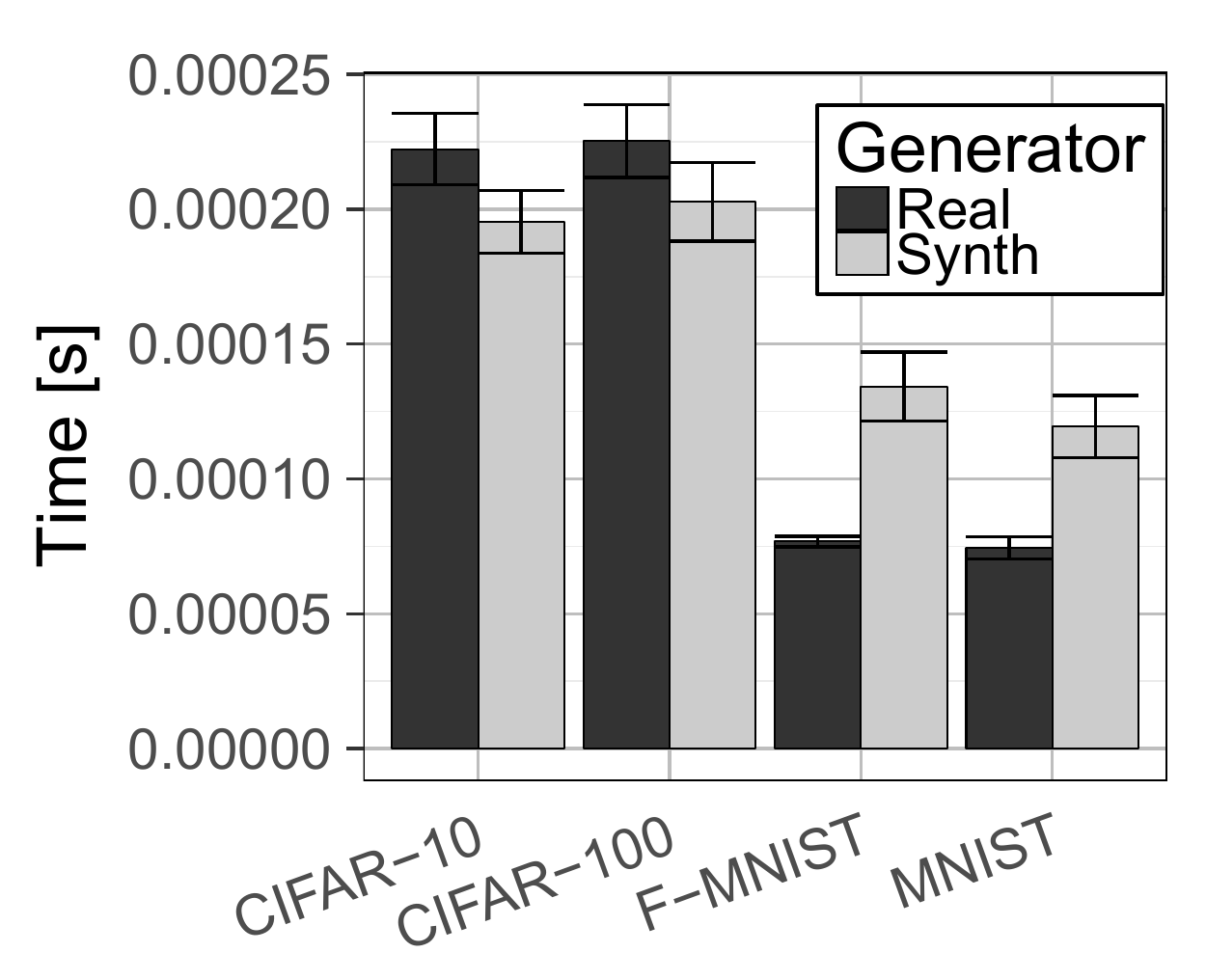}
	\includegraphics[width=0.5\linewidth]{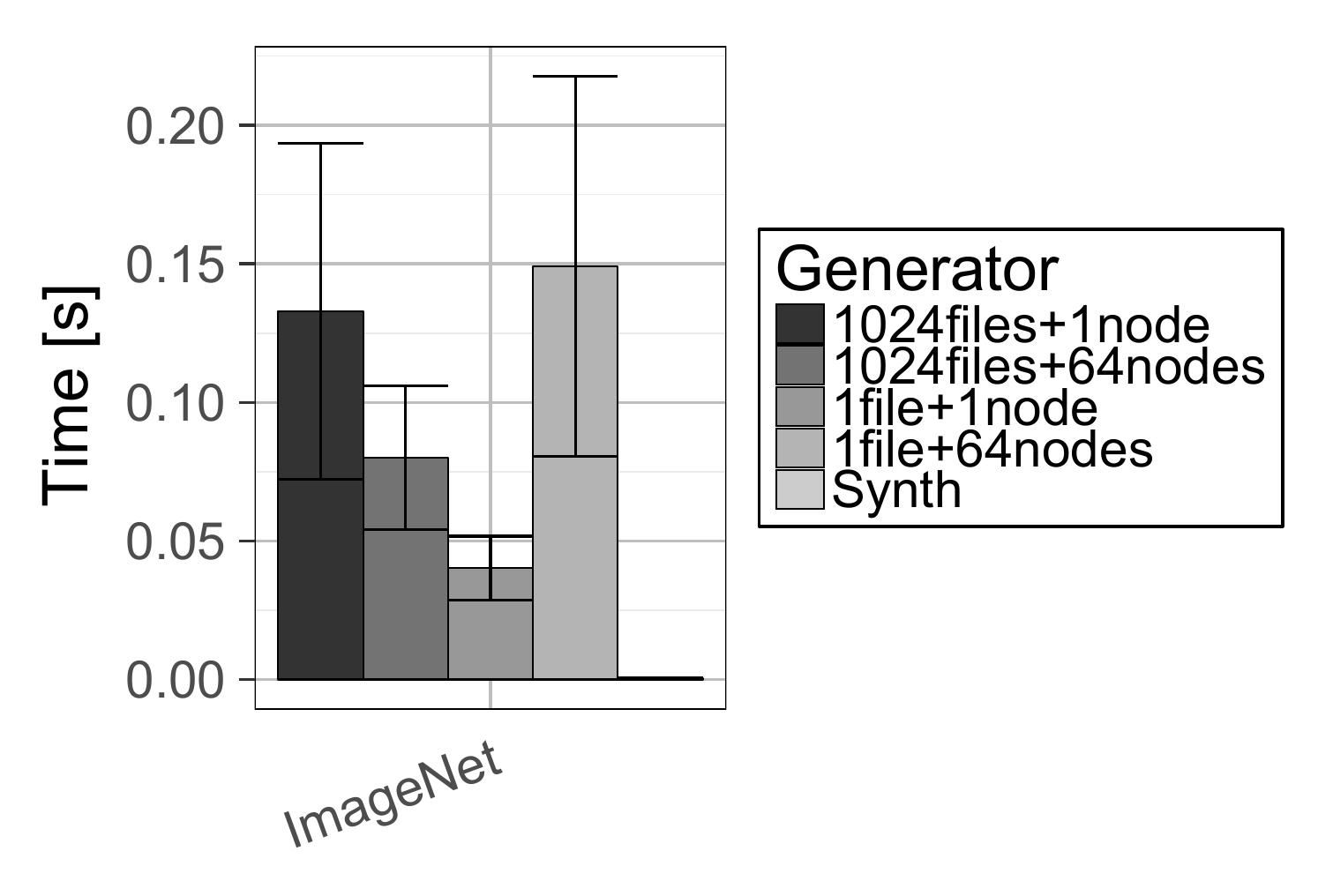}
	\vspace{-0.5em}
	\caption{The latency of loading various datasets.}
	\label{fig:latency}
	\vspace{-0.75em}
\end{figure}

\macs{Dataset Latency}
In Fig.~\ref{fig:latency}, we measure the latency of loading data from different image processing datasets (\texttt{DatasetLatency} metric), which include reading the files, decoding data, and constructing minibatches of size 128. We test different formats (binary, \texttt{TFRecord}, POSIX \texttt{tar}) on one node and a distributed training setting with 64 nodes. For each dataset, we measure both data loading as well as synthetic data generation. In the left-hand side of the figure (using raw binary files), we see that for small datasets (MNIST, Fashion-MNIST), data loading is faster than allocating and generating synthetic data. This is because the dataset is already stored in memory, and not encoded. In larger datasets, such as CIFAR-10 and 100, new data is occasionally loaded from the filesystem, thus synthetic generation is faster. 

For ImageNet, the figure (right-hand side) shows synthetic dataset generation is 2 orders of magnitude faster than loading images, as opposed to the other datasets. The main differences between ImageNet and the aforementioned datasets are the container (\texttt{TFRecord}) and image (encoded JPEG) format. To break down the image pipeline latency, we create an ImageNet dataset in a POSIX \texttt{tar} container with precomputed indexing (\texttt{IndexedTarDataset} in Deep500) and ingest it through two JPEG decoding pipelines (\texttt{PIL} and \texttt{libjpeg-turbo}). In Table \ref{tbl:imagenet_decoding} we see that using the \texttt{TFRecord} format and the TensorFlow pipeline is advantageous over the \texttt{tar} format. When using the POSIX format, both random file access and JPEG decoding play a role in slowdown. Even though for a single image \texttt{libjpeg-turbo} decodes images faster than the TensorFlow native decoder, the ratios between runtime of a minibatch and one image suggest that TensorFlow employs parallel decoding. Additionally, as opposed to the true random image selection in the \texttt{tar} format, TensorFlow uses pseudo-shuffling, where a buffer of (10,000) images is loaded into memory once and shuffled internally. This chunk-based loading reduces stochasticity, but, as the table shows, enables pipelining file I/O and in-memory shuffling.

\begin{table}[h]
\scriptsize
\sf
\centering
\begin{tabular}{lrrr}
	\toprule
	\textbf{Data Type} & \multicolumn{3}{c}{\textbf{Time [ms]}}\\\cline{2-4}\addlinespace
                       &	\multicolumn{2}{c}{\textbf{Indexed \texttt{tar}}} & \textbf{\texttt{TFRecord}} \\	\cline{2-3}\addlinespace
	                   & \texttt{PIL} & \texttt{libjpeg-turbo} & Native Decoder \\
	\midrule
	1 image (sequential)    & 10.04 & 2.80 & 7.43 \\
	1 image (shuffled)      & 51.19 & 34.60 & 9.50 \\
	128 images (sequential) & 1,378.34 & 315.18 & 127.02 \\
	128 images (shuffled)   & 6,849.45 & 6,433.72 & 139.13 \\
	\bottomrule
\end{tabular}
\caption{ImageNet decoding latency breakdown (median time).}
\label{tbl:imagenet_decoding}
\end{table}

For the distributed experiment, we test the ImageNet training set sharded to 1024 files (default) vs. 1 large file. In HPC, Parallel File Systems (PFS) generally prefer one segmented file rather than querying strings and inodes. Indeed, the latency of loading one file on a single node is lower than 1024. However, when using 64 nodes, we observe that surprisingly, 1024 files are $\approx$10\% faster on Piz Daint.
In all cases, the latency of loading a batch can be hidden by pipelining loading with DNN computation, a technique that is standard practice in large-scale ML.

\begin{figure}[h]
	\vspace{-1em}
	\centering
	\includegraphics[width=0.9\columnwidth]{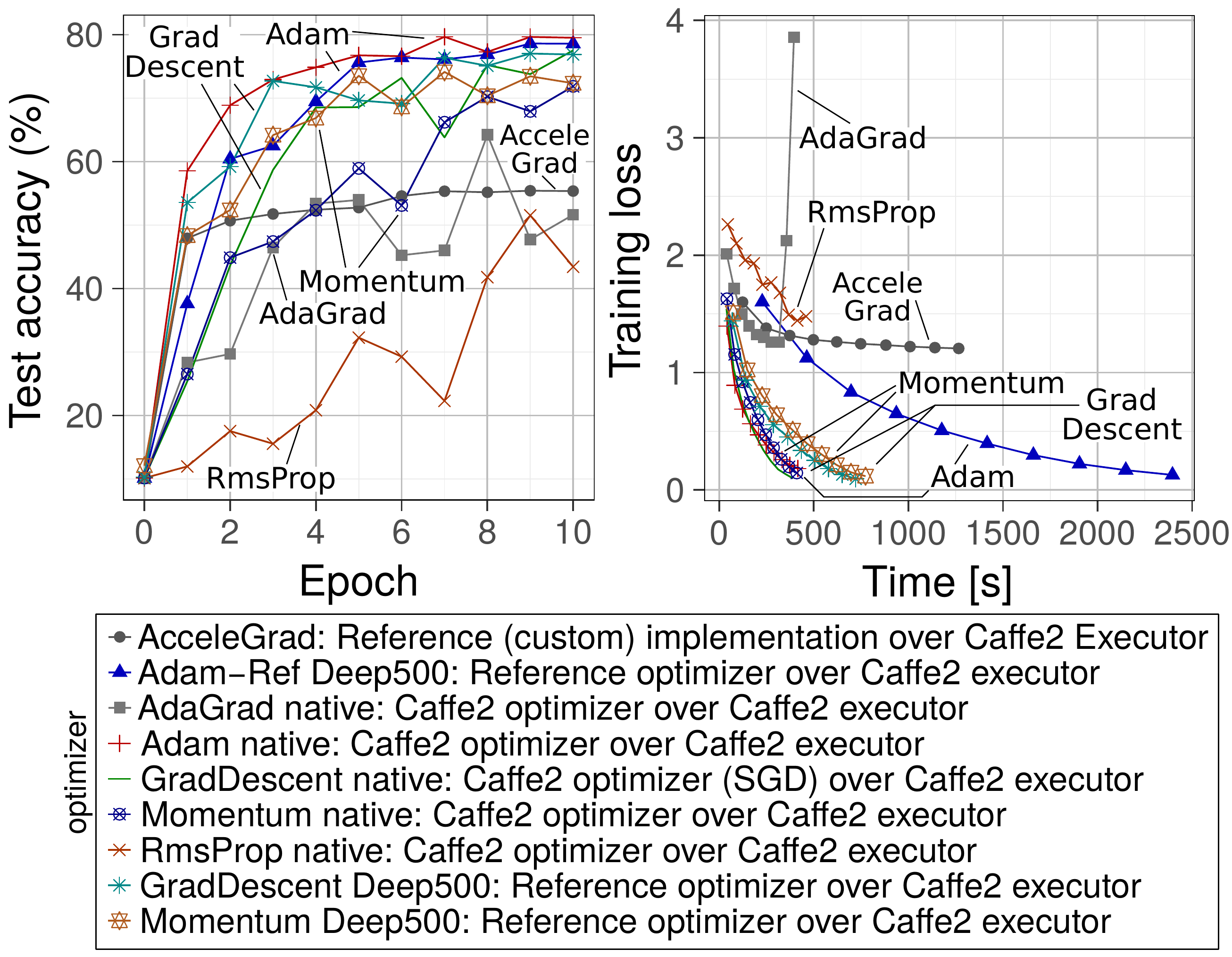}
	\vspace{-0.5em}
	\caption{The analysis of test accuracy vs.~epoch number and training loss
		vs.~elapsed time for different optimizers (assuming Caffe2, ResNet-18, CIFAR).}
	\label{fig:caffe2-optimizers}
\end{figure}

\begin{figure}[h]
	\vspace{-1em}
	\centering
	\includegraphics[width=0.9\columnwidth]{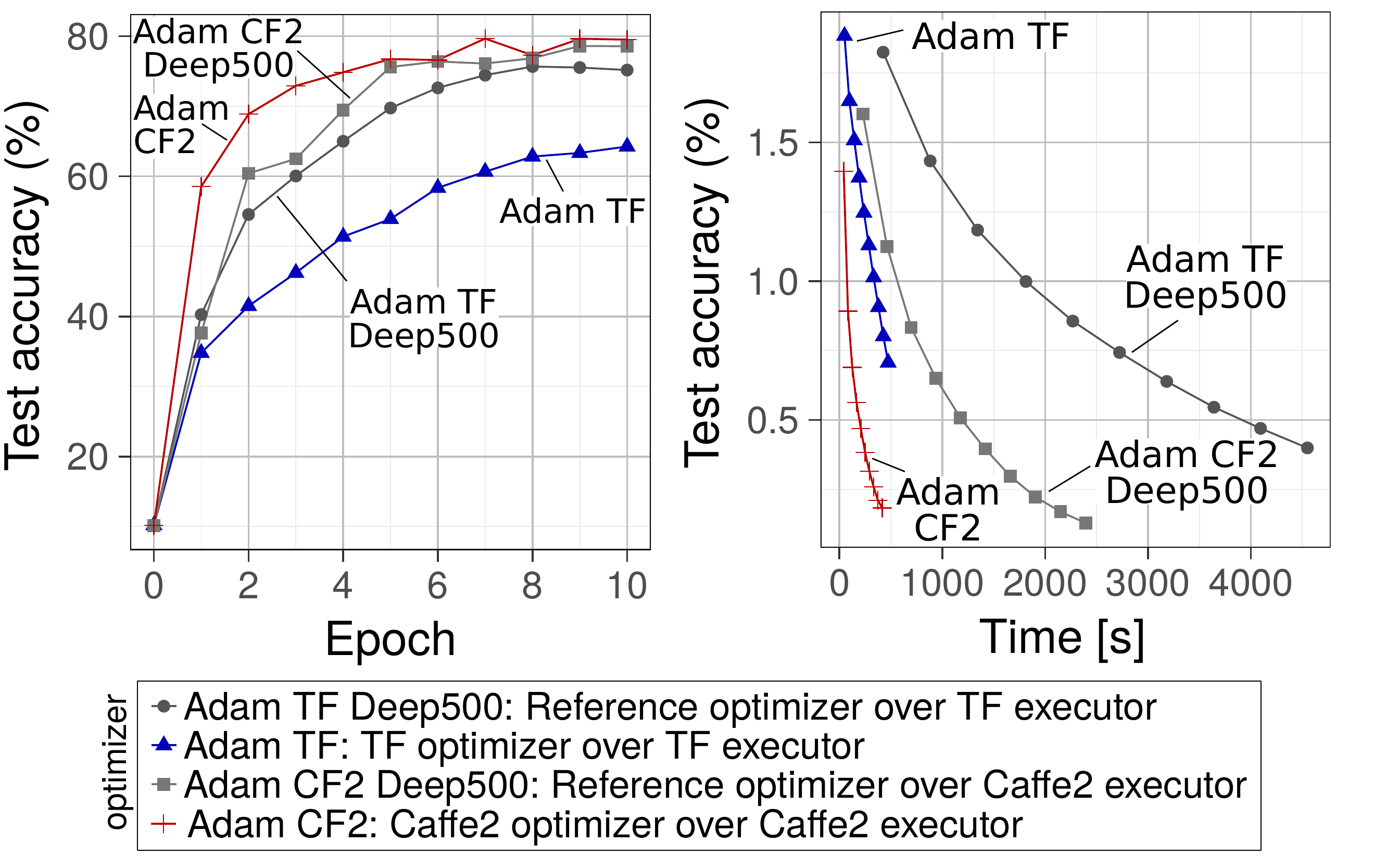}
	\vspace{-0.5em}
	\caption{The analysis of test accuracy vs.~epoch number and training loss
		vs.~elapsed time for different frameworks (assuming Adam, ResNet-18, CIFAR).}
	\label{fig:adam-frameworks}
\end{figure}

\macs{Convergence}
In Fig.~\ref{fig:caffe2-optimizers} and \ref{fig:adam-frameworks}, we analyze the \emph{convergence} of different native optimizers
in Caffe2, Deep500, and of
AcceleGrad~\cite{levy2018online}, the Deep500 custom Python operator from
Listing~\ref{lst:level-1-accelegrad}. While the AcceleGrad
algorithm is short and descriptive, it exhibits lower performance than the native Caffe2 optimizers ($\approx$1.6$\times$ slower). This can be attributed to the native Caffe2 weight update kernels, which are written specifically for GPUs. AcceleGrad also achieves comparable accuracy to similar algorithms (e.g., AdaGrad). 
Furthermore, while Deep500's Adam, which was directly translated from the
original algorithm~\cite{adam}, is $\approx$5$\times$ slower than native, it still achieves high accuracy, even when the framework does not (Fig.~\ref{fig:adam-frameworks}).

\begin{figure}[t]
	\vspace{-0.5em}
	\centering
	\begin{subfigure}{.49\columnwidth}
		\includegraphics[width=0.9\textwidth]{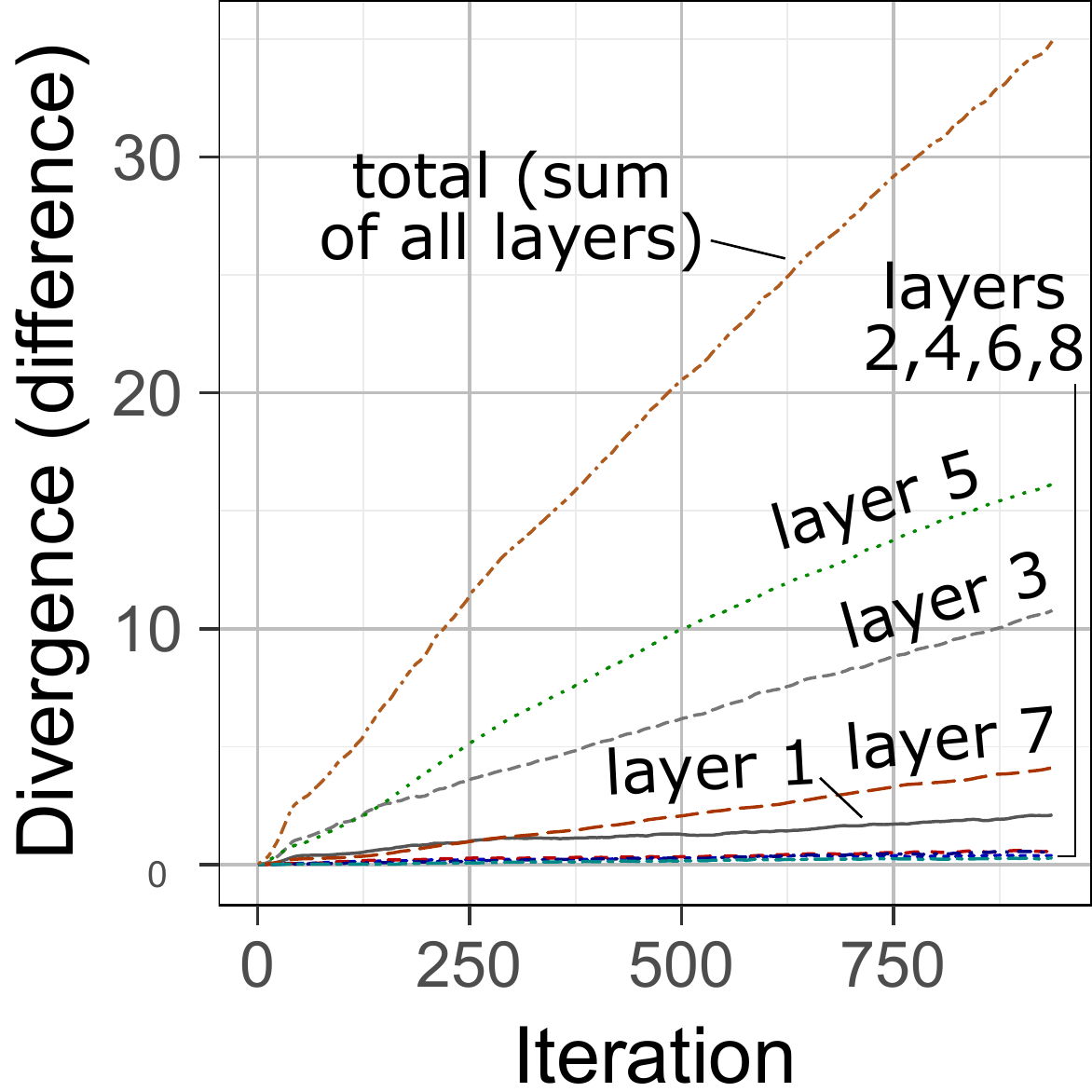}
		\caption{Divergence for the $\ell_2$ norm.}
	\end{subfigure}
	\begin{subfigure}{.49\columnwidth}
		\includegraphics[width=0.9\textwidth]{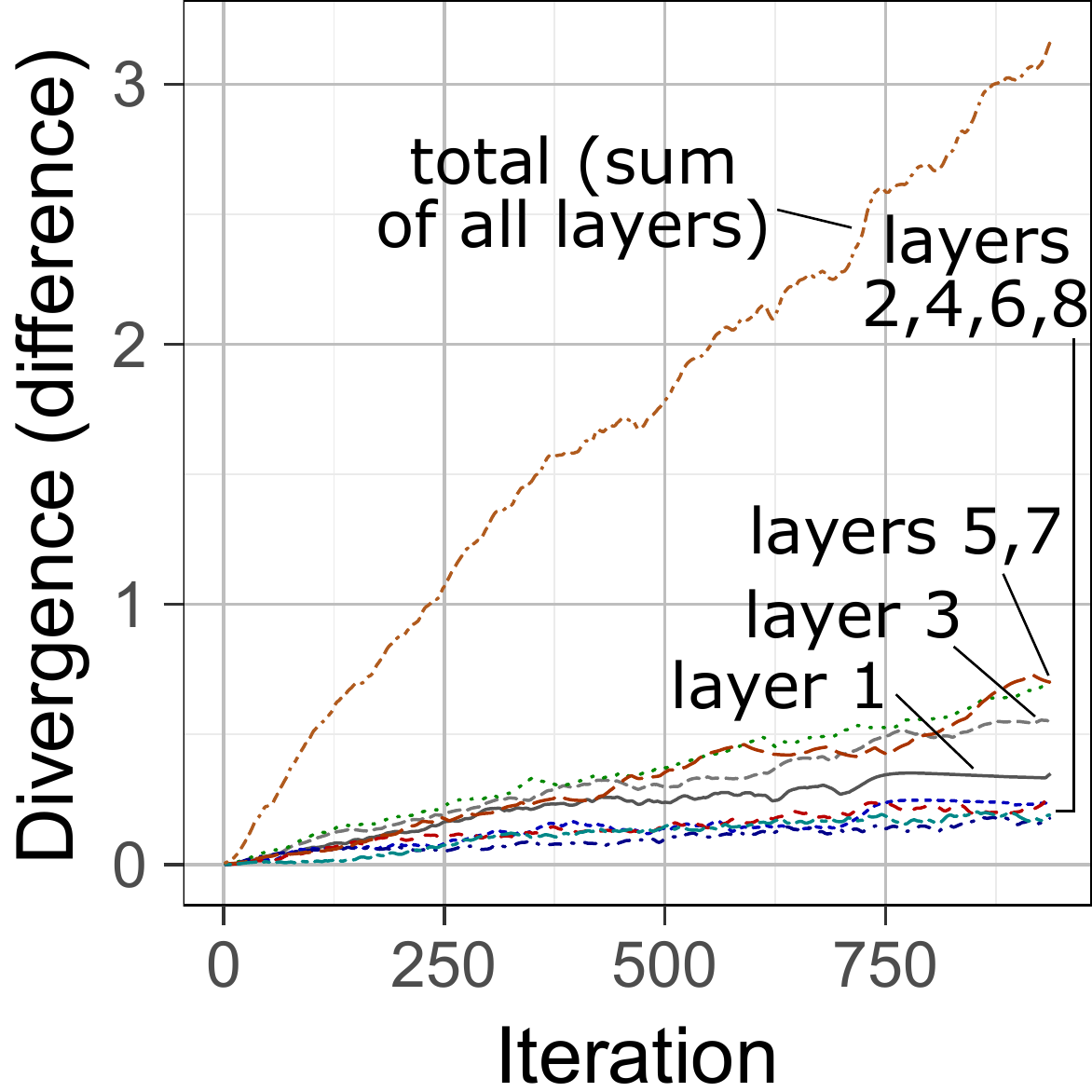}
		\caption{Divergence for the $\ell_{\infty}$ norm.}
	\end{subfigure}
	
	\vspace{-0.25em}
	\caption{The difference (divergence) between DNN weights in the native optimization (Adam on TensorFlow) and the Deep500 Adam
		optimization (MNIST).}
	\vspace{-0.25em}
	\label{fig:adam-tf-divergence}
\end{figure}

In an attempt to understand this difference, we test the accuracy of the TensorFlow Adam optimizer on a smaller scale by comparing its trajectory with the Deep500 implementation. Fig.~\ref{fig:adam-tf-divergence} shows the $\ell_2$ and $\ell_\infty$ norms of the difference between the
parameters, illustrating the chaotic divergence of deep learning, now easily
visualized by Deep500. We observe that a single step of
TensorFlow is faithful to the original algorithm, however, continuing training increases divergence, where some parameters (e.g., fully connected, layers 5,7) diverge faster than others (additive bias, layers 2,4,6,8).

The Deep500 reference optimizers are evidently slower (e.g., Figure~\ref{fig:adam-frameworks}), as they are unoptimized reference implementations. Here, our primary goal is \emph{not}
to offer tuned competitive codes, but instead to illustrate that Deep500 enables
comparison and convergence of a multitude of optimizers.

\subsection{Level 3: Distributed Training}

\begin{figure*}[t]
	\vspace{-1em}
	\centering
		\includegraphics[height=2in]{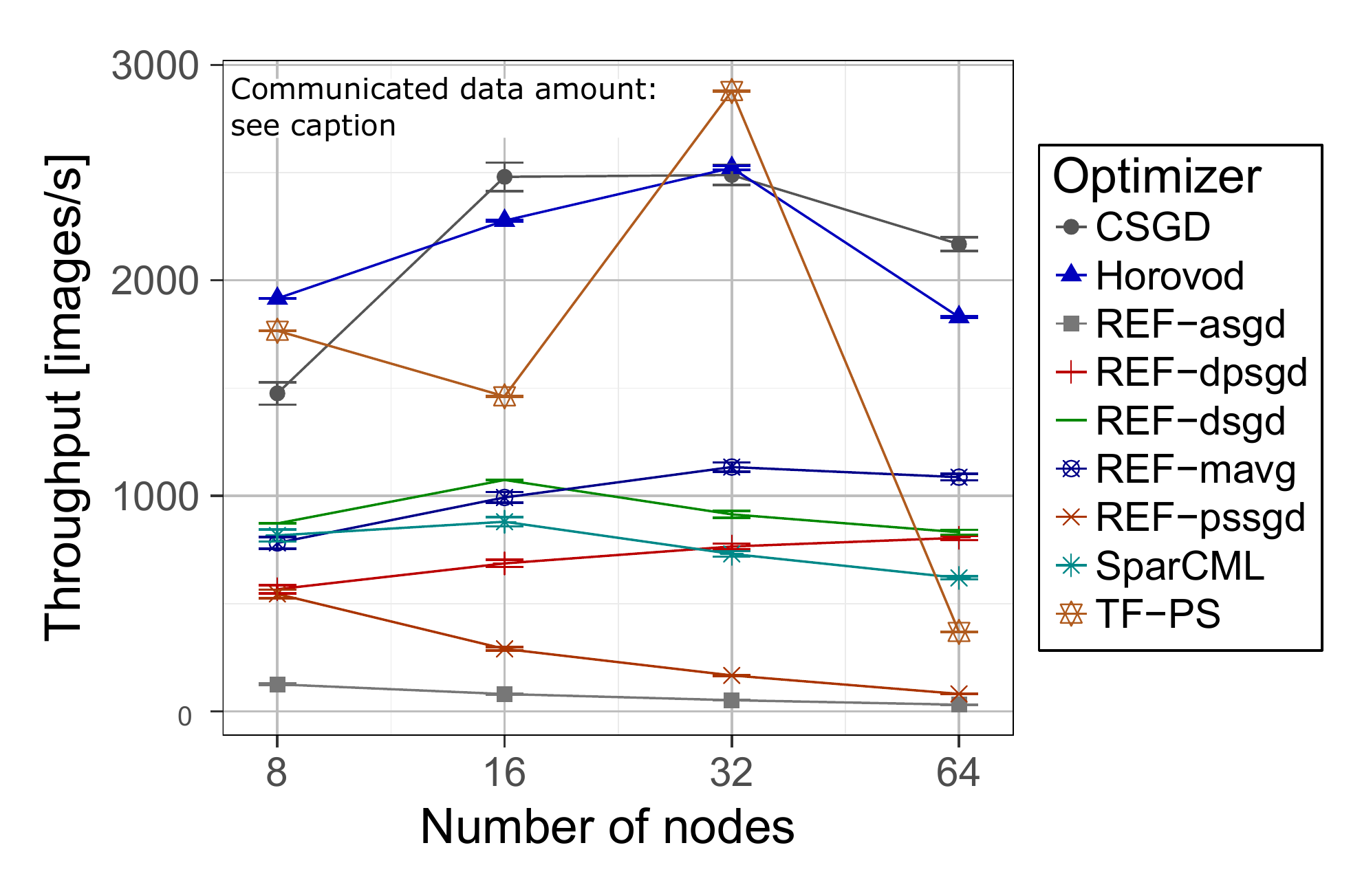}
		\includegraphics[height=2in]{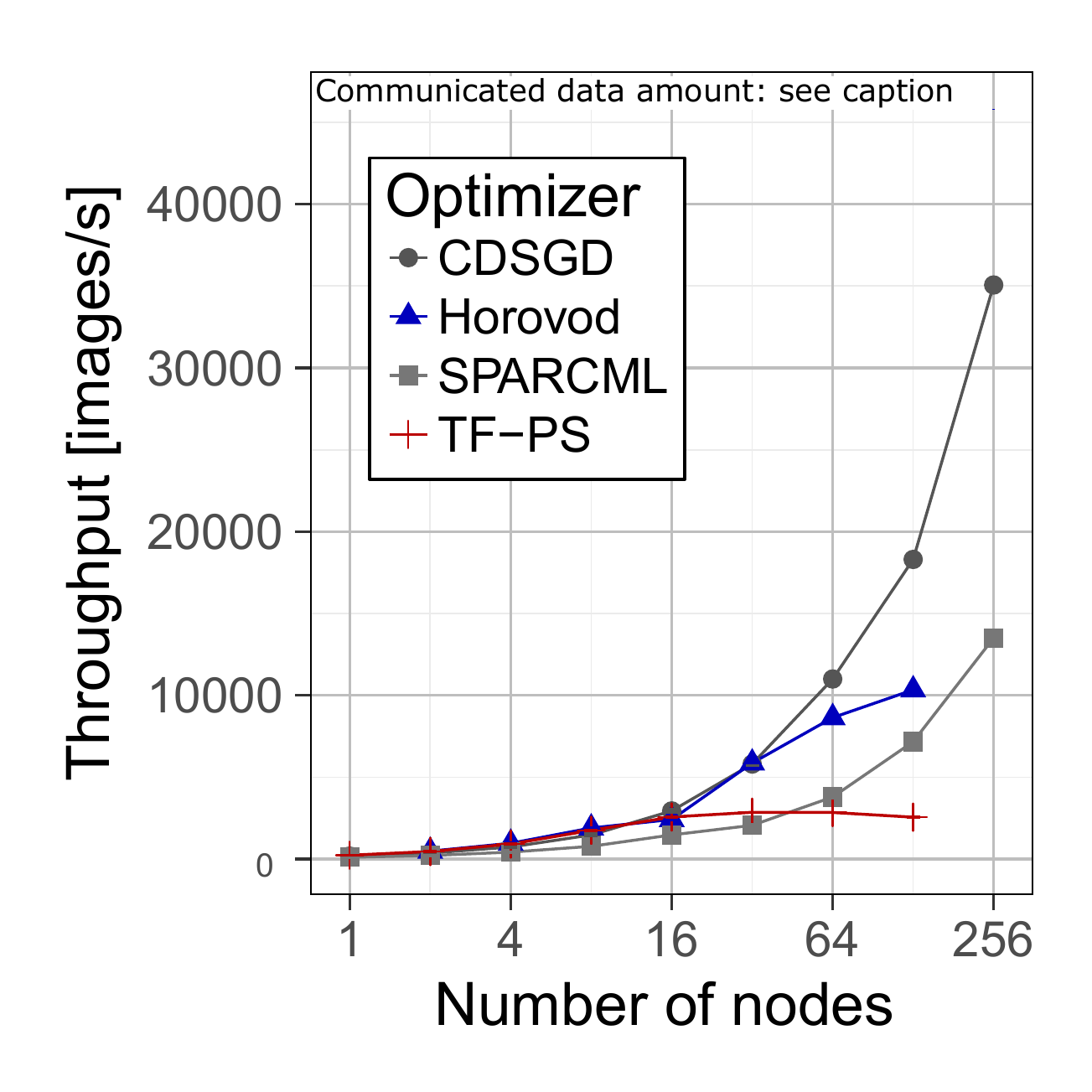}
	\vspace{-0.25em}
	\caption{\textbf{Scaling of Level~3}: Strong (left) and weak (right) scaling on Piz Daint and ImageNet.
  \textbf{Communicated data per node:} 0.952 GB (CDSGD), 0.951 GB (SparCML), 
0.952 GB (REF-dsgd), 
28.573 GB (REF-asgd),
1.904 GB (REF-dpsgd),
1.903 GB (REF-pssgd).
}\label{fig:scaling}
\end{figure*}

Finally, we analyze Deep500's Level~3. \emph{We compare distributed variants of
SGD}, including TensorFlow's native parameter server (TF-PS), Horovod, as well as Deep500 reference implementations of centralized SGD (PSSGD), decentralized (DSGD), decentralized with a neighbor-based communication graph (DPSGD), asynchronous (ASGD), model-averaging (MAVG), DSGD with a Deep500 custom C++/MPI \textit{allreduce} operator (CDSGD), and the \textit{custom
distributed communication} scheme SparCML~\cite{renggli2018sparcml}, written
as a custom Deep500 operator. All compared Deep500 implementations are distributed over the TensorFlow graph executor. We use 
ResNet-50 for strong and weak scaling, and up to 256 compute
nodes. We use the \texttt{CommunicationVolume} metric in conjunction with mpiP~\cite{mpip} to collect communication statistics.

Fig.~\ref{fig:scaling} (left) presents strong scaling results of the distributed implementations and competitors, compared with two baselines --- TF-PS and Horovod --- both measured using the TensorFlow Benchmark\footnote{https://www.github.com/tensorflow/benchmarks}. We use a minibatch size of 1,024 images on
8--64 nodes (since fewer nodes run out of memory and more nodes become
ineffective). The figure shows that while \textit{Python distributed optimizers
provide reference results for \textbf{correctness}} analysis, \textit{C++
operators can deliver \textbf{high-performance}} necessary for large-scale
training, which are on-par with the state-of-the-art (Horovod). In particular, several effects can be seen: \ding{182} ASGD is
centralized but does not use broadcast/gather operations. Consequently, 
despite being asynchronous, ASGD becomes
slower the more worker nodes queue up to communicate. \ding{183}~PSSGD, MAVG, and DSGD
all start with similar epoch times, but as nodes increase, the decentralized
versions (MAVG, DSGD) prove more efficient. \ding{184} DSGD written in C++, which uses
direct CPU/GPU pointers, scales strongly up to 32 nodes and is almost an order
of magnitude faster than its Python counterpart, which undergoes conversions
to/from NumPy arrays.

In terms of communication volume, the collected metrics indicate that the reference DSGD and C++ DSGD exhibit the same communication volume, as expected. The number of PSSGD messages, however, scales linearly with the number of nodes. DPSGD communication volume remains constant with respect to the number of nodes, but usually converges slower and to a less accurate result~\cite{ben2018demystifying}. As for SparCML's sparse allreduce (183 lines of C++ code), we see that while
communication is greatly reduced (up to 2$\times$ on 8 nodes), the running
time is still high compared to the DSGD \textit{allreduce} custom operator (23
lines of code), and increases with the number of participating nodes. This is
both due to the reduced vector representation, which becomes denser with increasing nodes~\cite{renggli2018sparcml} (every allreduce step aggregates more sparse vectors with different indices), and
due to the time it takes to filter the dense gradient to the sparse
representation, which could potentially be optimized by using a CUDA custom
operator.

In Fig.~\ref{fig:scaling} (right), we study the weak scaling of the same implementations on 1--256 nodes. The baseline for this test is TF-PS, the default distributed implementation available in TensorFlow. Although simple, the \textit{allreduce} operator (CDSGD) provides full
DSGD, and is able to scale better than the PS architecture and Horovod, as in point
\ding{183} above. This result is also non-trivial, since the parameters are
stored on the GPUs, and they are copied automatically using Deep500 for use
with MPI. Also observe that the native TensorFlow and Horovod implementations are missing results at 256 nodes. For TF-PS, the application crashed, whereas on Horovod the test ran but produced exploding (infinitely increasing) loss values, which is an indicator of incorrect gradient accumulation.

Overall, the plots show that \textbf{using Deep500, comparing multiple
communication schemes is as easy as replacing an operator}. Deep500 facilitates
the tradeoff analysis between different topologies, operator overhead (e.g.,
gradient sparsification), and optimizer quality (async. vs. sync. SGD); and
enables benchmarking results on large node configurations.

\section{Related Work}

Our work touches on various areas. We now discuss related works, briefly
summarizing the ones covered in previous sections: DL frameworks
in~\cref{sec:back-fr} and Table~\ref{tab:intro-frameworks}, DL data model and
format in~\cref{sec:back-onnx}, and DL benchmarks in
Table~\ref{tab:intro-benchmarks}.
 
\macs{DL Benchmarks}
The DL community has recently gained interest in benchmarking DL codes.
Example benchmarks are DAWNBench~\cite{coleman2017dawnbench},
MLPerf~\cite{mlp}, or DeepBench~\cite{deepbench}; see
Table~\ref{tab:intro-benchmarks} for a full list and analysis of their
functionalities.
\emph{Deep500 is the only benchmark that addresses \textbf{the five challenges}
described in~\cref{sec:challenges}: customizability, metrics, performance,
validation, and reproducibility}.

\macs{DL Frameworks}
There exist many DL frameworks and related libraries as well as
frontends~\cite{abadi2016tensorflow, jia2014caffe, chen2015mxnet,
paszke2017automatic}.  As we illustrate in Table~\ref{tab:intro-frameworks},
none of them offers a full spectrum of functionalities.
\emph{Deep500 does not only enable benchmarking of these systems. On top of
that -- through its meta-framework design -- it enables integrating arbitrary
elements of the considered DL systems to \textbf{combine the best of different
DL worlds}}.

\macs{DL Data Formats}
Finally, we use and extend the established ONNX DNN
format~\cite{exchange2018onnx} with an object-oriented notation, new
operations, and others.
Thus, \emph{Deep500 significantly improves interoperability between ONNX and DL
frameworks}.

\section{Conclusion}

Deep Learning (DL) has become ubiquitous in areas as diverse as speech
recognition and autonomous driving. However, it is still unclear how to compare
and benchmark the plethora of available DL algorithms and systems, especially
in extreme-scale distributed environments.

To answer these questions, we introduce Deep500: a customizable infrastructure that enables
detailed, accurate, fast, and fair benchmarking of DL codes. The essence of
Deep500 is its \emph{layered and modular} design that allows to \emph{independently} extend
and benchmark DL procedures related to simple operators, whole neural networks,
training schemes, and distributed training. The principles behind this 
design can be reused to enable interpretable and
reproducible benchmarking of extreme-scale codes in domains outside DL.

To ensure the best design decisions for Deep500, we analyze challenges in
benchmarking complex and large-scale DL codes. To this end, we identify \emph{five
core challenges}: customizability, metrics, performance, validation, and reproducibility.  Through extensive evaluation, we
illustrate that Deep500 satisfies these challenges.  For example, it ensures
identical accuracy while offering negligible ($<$1\%) performance overheads
over native operators or whole DNNs in state-of-the-art DL frameworks,
including TensorFlow, PyTorch, and Caffe2.

Finally, we predict that Deep Learning will become a part of computing as
important as general dense or sparse linear algebra.
Thus, we construct Deep500 such that it can be freely modified to ensure fair benchmarking, produce artifacts for DL papers, provide insightful analyses, and enable effective development of any future DL effort.

\section*{Acknowledgments}

We thank C\'{e}dric Renggli, Dan Alistarh, Yosuke Oyama, and Kfir Y. Levy for valuable discussions; 
and Hussein Harake, Colin McMurtrie, and the whole CSCS team granting
access to the Greina and Daint machines, and for their excellent technical
support.
This project has received funding from the European Research Council (ERC) under the European Union's Horizon 2020 programme (grant agreement DAPP, No. 678880).
T.B.N. is supported by the ETH Zurich Postdoctoral Fellowship and Marie Curie Actions for People COFUND program. 

\bibliographystyle{IEEEtran}
\bibliography{IEEEabrv,references}

\end{document}